\documentclass[prb,a4paper,twocolumn,showpacs,floatfix]{revtex4-1}
\usepackage[utf8]{inputenc}
\usepackage[T1]{fontenc}
\usepackage{lmodern}
\usepackage{amssymb}
\usepackage{amsmath}
\usepackage{amsbsy}
\usepackage{bm}
\usepackage{esint}
\usepackage{graphicx}
\usepackage{color}
    \newcommand{\I}{\mathrm{i}}
    \newcommand{\E}{\mathrm{e}}

\begin{document}
\title{Nonlinear electric transport in graphene with magnetic disorder}
\author{Arnaud Demion}
\author{Alberto D.\ Verga}\email{Alberto.Verga@univ-amu.fr}
\affiliation{Université d'Aix-Marseille, IM2NP-CNRS, Campus St.\ Jérôme, Case 142, 13397 Marseille, France}
\date{\today}
\begin{abstract}
   The influence of magnetic impurities on the transport properties of graphene is investigated in the regime of strong applied electric fields. As a result of electron-hole pair creation, the response becomes nonlinear and dependent on the magnetic polarization. In the paramagnetic phase, time reversal symmetry is statistically preserved, and transport properties are similar to the clean case. At variance, in the antiferromagnetic phase, the system undergoes a transition between a superdiffusive to a subdiffusive spreading of a wave packet, signaling  the development of localized states. This critical regime is characterized by the appearance of electronic states with a multifractal geometry near the gap. The local density of states concentrates in large patches having a definite charge-spin correlation. In this state, the conductivity tends to half the minimum conductivity of clean graphene.
\end{abstract}
\pacs{72.80.Vp,72.15.Rn}
\maketitle
\section{Introduction}
\label{s:intro}

Electronic transport in graphene exhibits unique properties that stem from the nature of its quasiparticles, two-dimensional Dirac fermions \cite{Novoselov-2005kx,Ando-2008rr,Castro-Neto-2009fk,Das-Sarma-2011fk}. Under the action of a weak static electric field, the linear response theory predicts a minimal conductivity, characteristic of the linear dispersion near the Dirac point \cite{Ludwig-1994fk}. This conductivity is insensitive to weak disorder, as a consequence of the absence of Anderson localization when intervalley scattering can be neglected \cite{Bardarson-2007zr,Nomura-2007fk}. Increasing the disorder paradoxically facilitates the conduction through the Klein tunneling mechanism \cite{Beenakker-2008vn}. Short range disorder that allows intervalley transitions \cite{Lherbier-2008fk}, spin dependent scattering triggered for instance by magnetic impurities \cite{Withoff-1990ly,Hu-2011fk}, and electric potential differences as in \(n\)-\(p\) junctions that introduce nonlinear corrections to the conductivity \cite{Cheianov-2006uq}, can qualitatively change the transport properties of pristine graphene. In particular, under a strong electric field a new phenomenon arises, the Schwinger electron-hole pair production \cite{Schwinger-1951fk,Allor-2008fk}. It has been demonstrated experimentally that pair creation modifies the current-voltage characteristics \cite{Vandecasteele-2010uq}. A power law was found with a mobility dependent exponent taking values between the linear response and the pair production dominated response (exponent \(3/2\)).

Transport properties are related to the electronic band structure of graphene, which can be modified by various mechanisms including the scattering off impurities or vacancies \cite{Peres-2010uq}, and by perturbations originating from random edge configurations \cite{Areshkin-2006fk,Rozhkov-2011fk}. These processes change the energy bands by populating the levels in the neighborhood of the Dirac point, and by changing their localization properties. In particular, doping graphene with magnetic impurities breaks the sublattice symmetry and opens a gap \cite{Dugaev-2006fk,Brey-2007kq,Yazyev-2010uq}. 

%
% Fig 1
\begin{figure}
  \centering
  \includegraphics[width=0.44\textwidth]{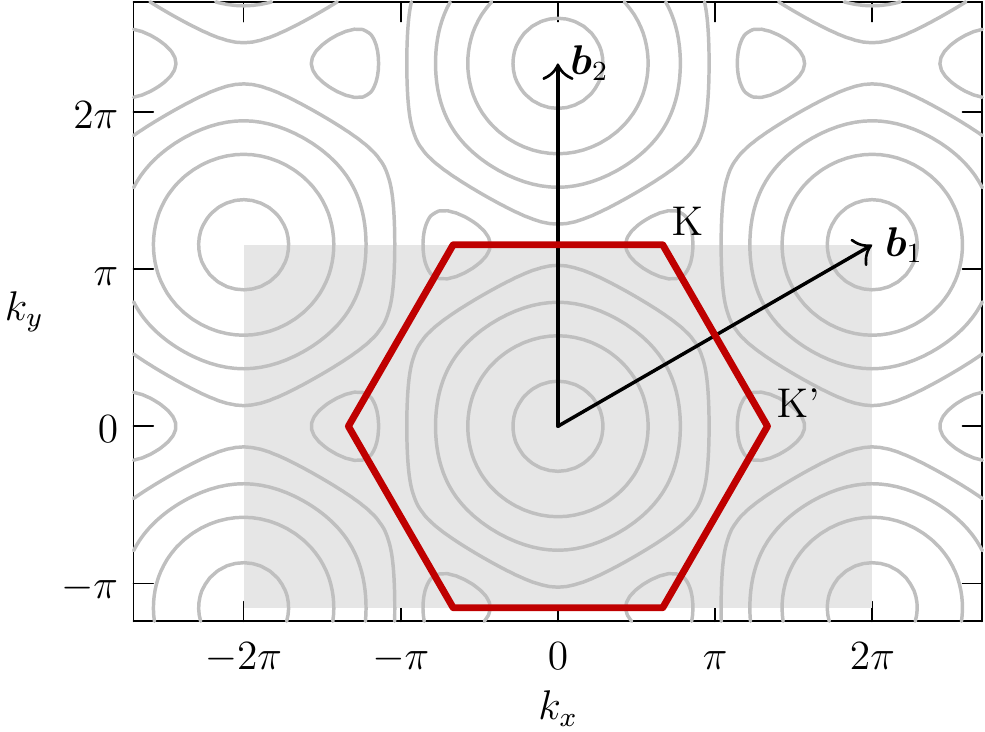}
  \caption{\label{f:1} (color online). Contours of the energy \(\epsilon_{\bm k}\) and Brillouin zone for pristine graphene (\(\mathit{BZ}\) red hexagon); \(\bm b_1\) and \(\bm b_2\) are the primitive vectors of the reciprocal lattice. The Dirac cones are located at K and K'. The light gray rectangular box defines the integration domain used in the numerical computations, it covers two cells.}
\end{figure}

In this paper we investigate the effect on the electronic transport of magnetic disorder in the strong electric field regime. We are interested in the dependency of the pair production rate and electric current on the intensity of the applied electric field. It is expected that under paramagnetic disorder the general picture of nonlinear transport is preserved \cite{Dora-2010kx,Kao-2010fk}, but that under magnetic order, this picture would change essentially as a consequence of localization \cite{Rappoport-2011vn}. In addition to the appearance of localized states, the opening of a gap induced by magnetically polarized impurities (magnetic state with spatial disorder), should significantly fade away the pair production, and consequently change the current-voltage characteristic.

%
% FIG 2
\begin{figure*}
\centering
\includegraphics[width=0.16\textwidth]{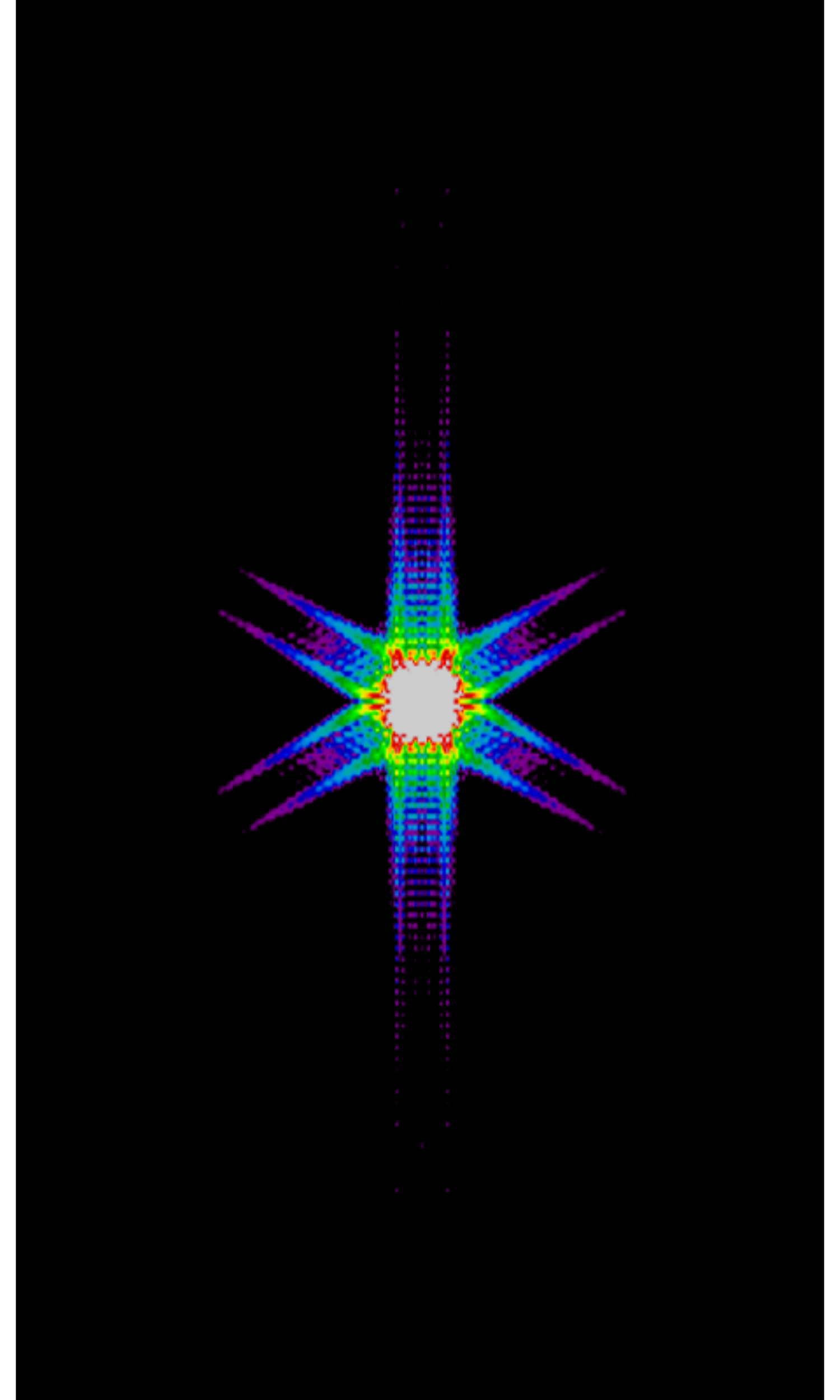}%
\includegraphics[width=0.16\textwidth]{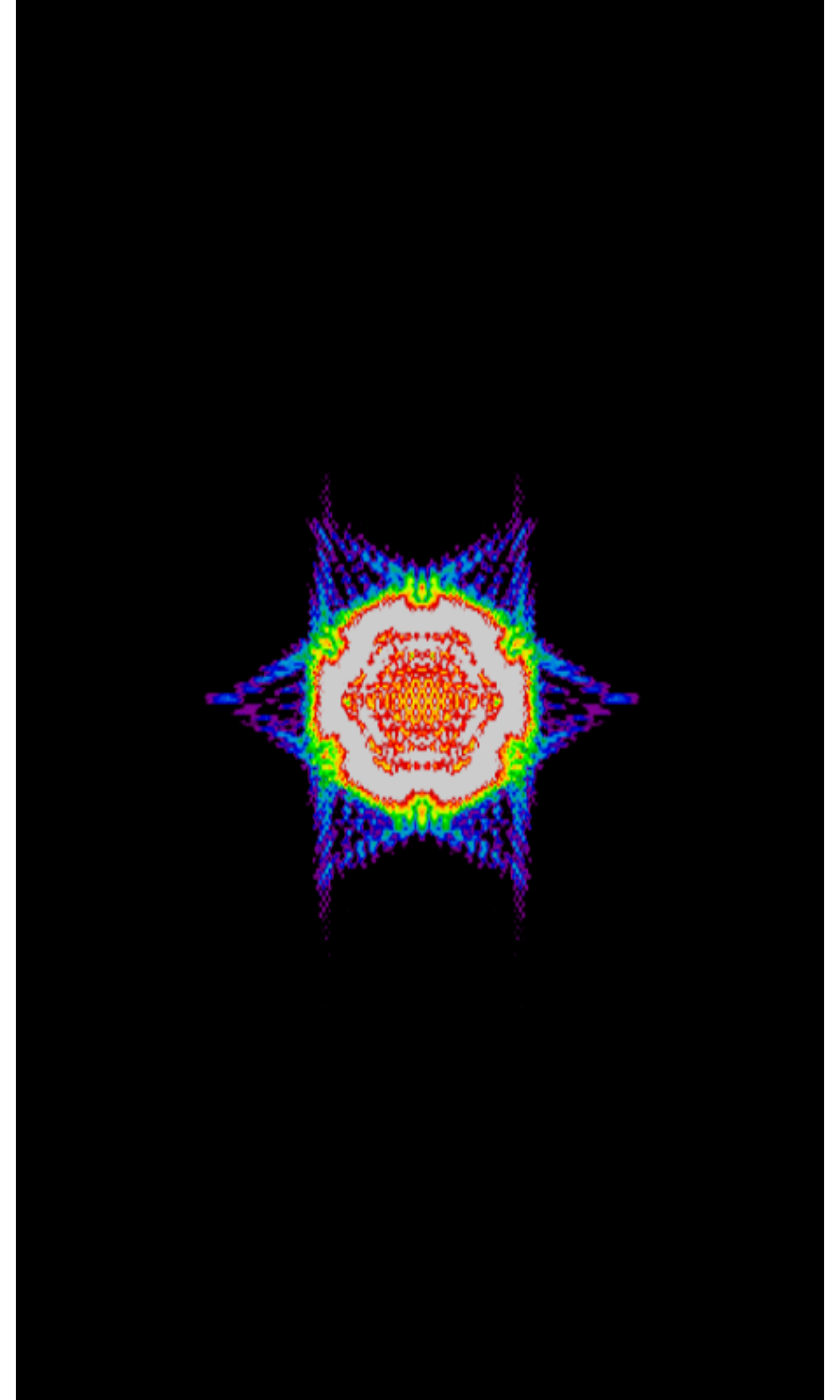}%
\includegraphics[width=0.16\textwidth]{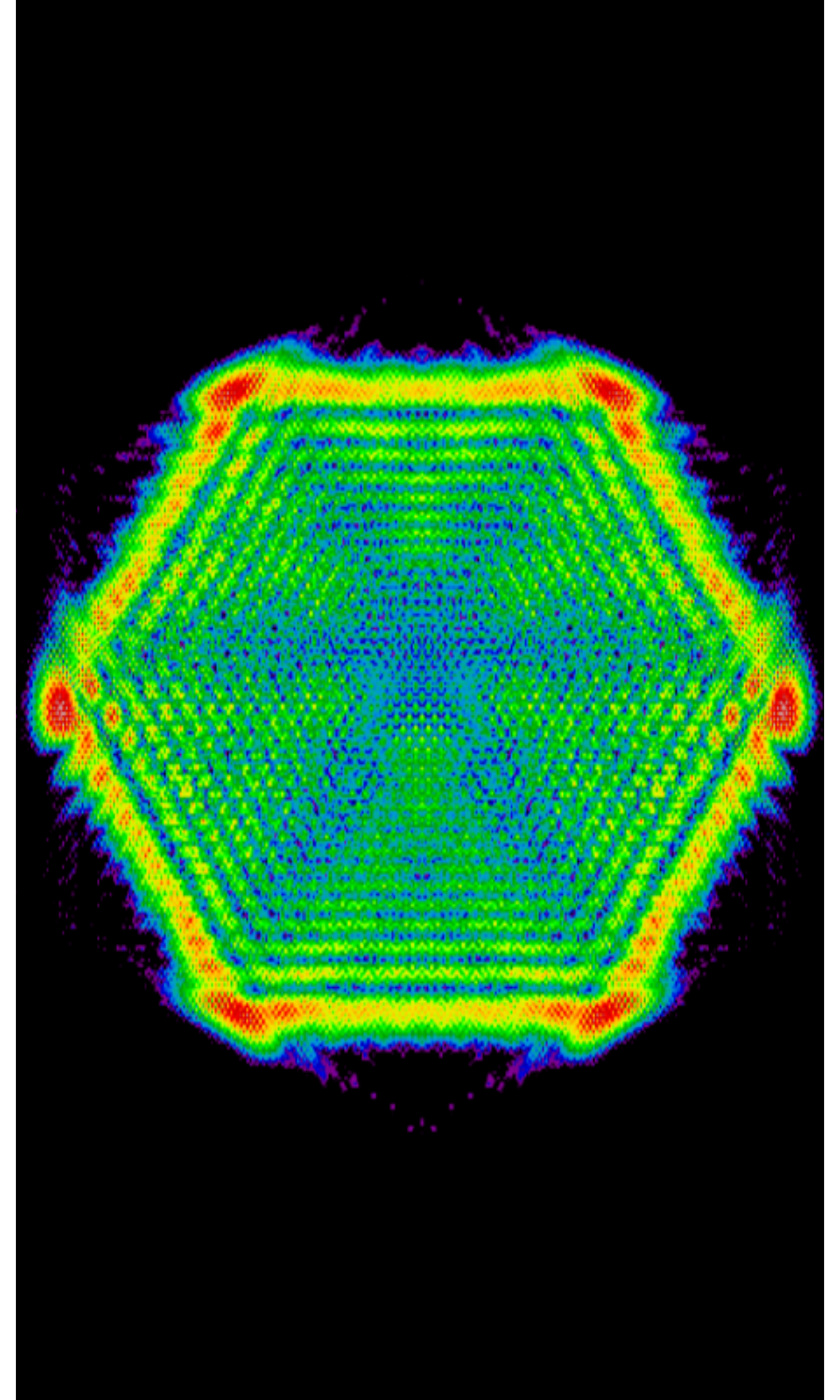}\hfill%
\includegraphics[width=0.16\textwidth]{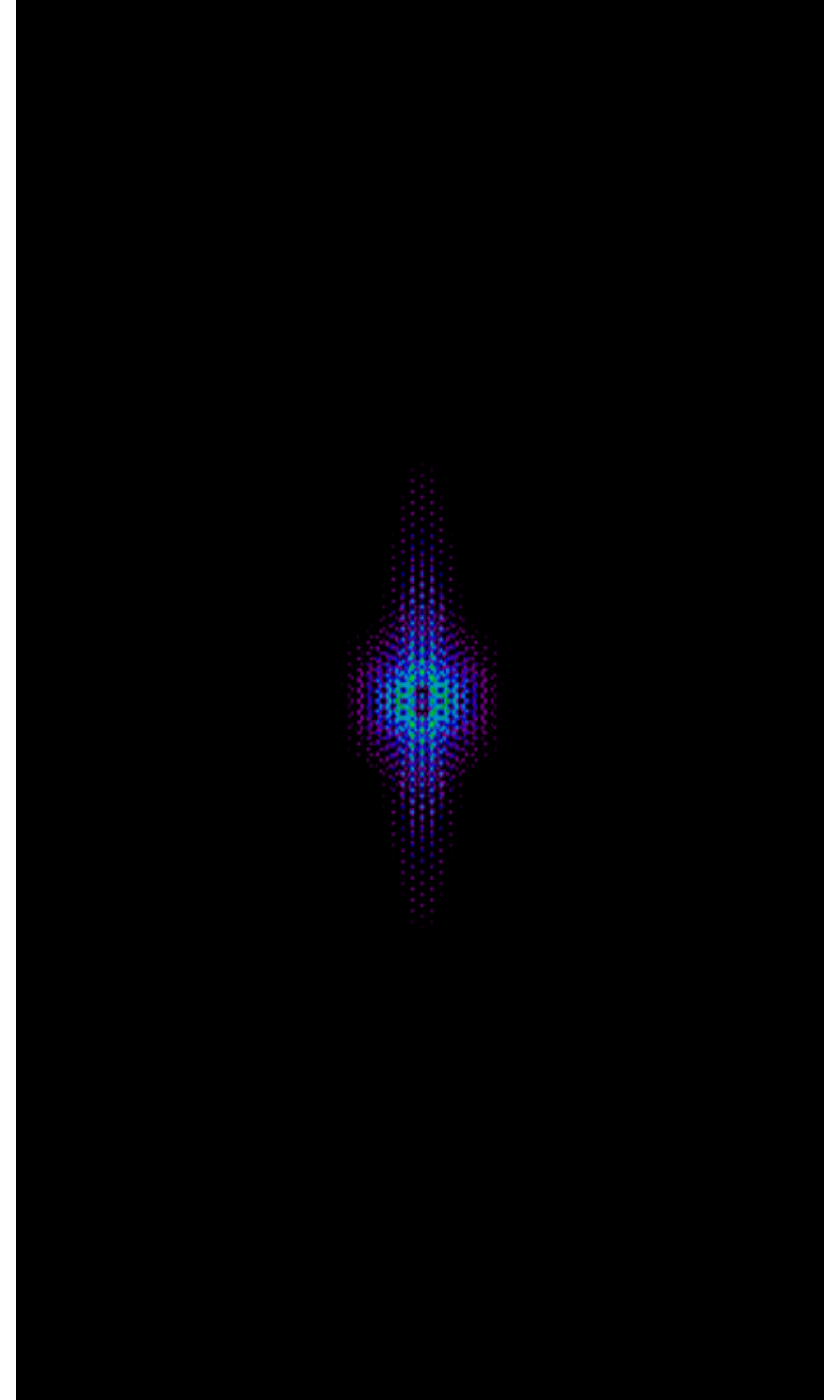}%
\includegraphics[width=0.16\textwidth]{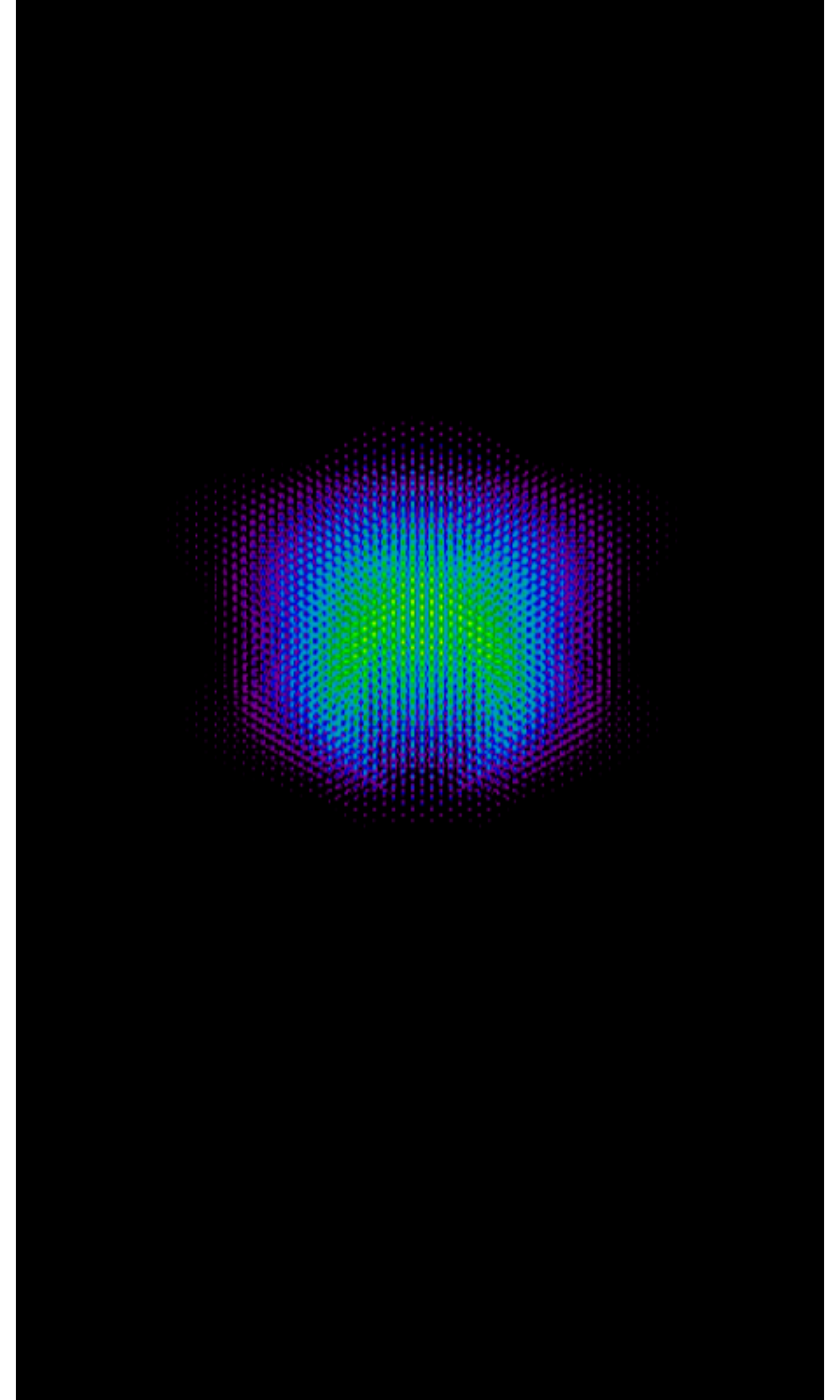}%
\includegraphics[width=0.16\textwidth]{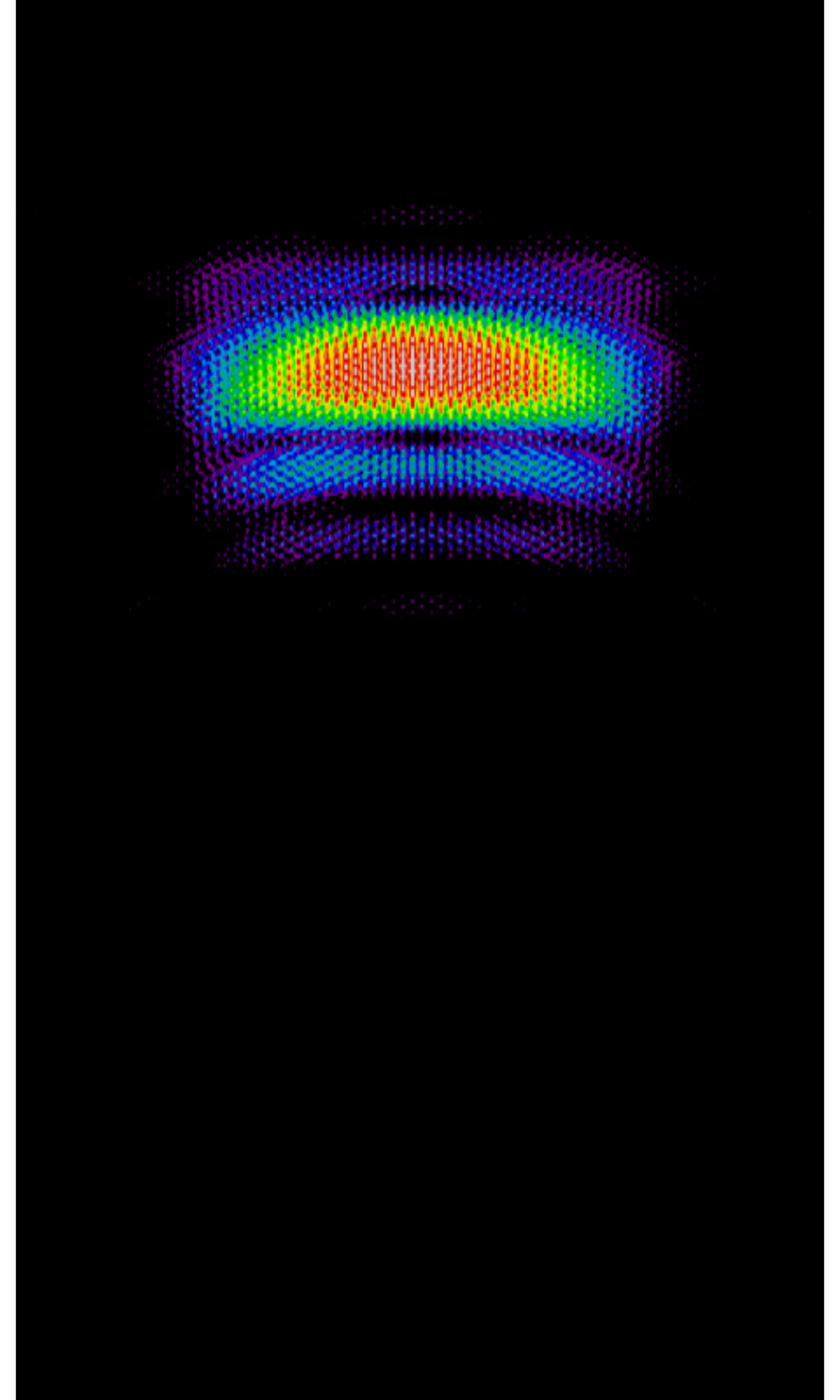}
\caption{\label{f:2} (color online). Temporal evolution of the hole (left) and electron (right) densities. Times \(t=4,16,60\,t_0\), electric field \(E_y=0.01\,E_0\), lattice \((2\times 256\, a)^2\) sites. The colormap (from blue to red) is in a logarithmic scale to enhance the small values of the wave function.}
\end{figure*}

We consider a tight-binding model where the coupling with randomly distributed magnetic moments is ensured by a simple exchange term \cite{Jacob-2010yu,Uchoa-2011fk}. The external electric field is derived from a vector potential. This allows us to minimize finite size effects by using periodic boundary conditions, and integrating the time-dependent Schrödinger equation in momentum space. The transport properties are studied by direct computation of the mean current and pair creation rate from the evolution of the wave function. We finally discuss the localization of electronic states using the local density of states as order parameter \cite{Dobrosavljevic-2003kx,Schubert-2010kx}. The numerical calculation of the density of states is performed using Chebyshev polynomials \cite{Weisse-2006fk}.

\section{Model of graphene in a strong electric field}
\label{s:model}

We describe electrons in graphene subject to an external electric field, by a two-dimensional tight-binding model with first neighbor interactions and randomly distributed classical magnetic impurities. We consider a hexagonal lattice with \(N\) sites, area \(L^2\), and constant \(a\), with two atoms \(A,B\) per cell. The primitive vectors are \cite{Ando-2008rr},
\begin{align}
    \bm a_1 &= a (1,0)\,, \\
    \bm a_2 &= a (-1/2,\sqrt{3}/2)\,,
\end{align}
and the reciprocal vectors,
\begin{align}
    \bm b_1 &= (2\pi/a) (1,1/\sqrt{3})\,, \\
    \bm b_2 &= (2\pi/a)(0,2/\sqrt{3})\,,
\end{align}
as can be seen in Fig.~\ref{f:1}. Let \(i\) be a lattice point of coordinates \(\bm x_i = (x_i,y_i)\) (\(i = 1, \ldots, N\)); the neighbors \(\bm x_j\) of \(\bm x_i\) are given by the three vectors \(\bm x_j = \bm x_i + \bm d_{ij}\), where \(\bm d_{ij}=\bm d_a\), (\(a=1,2,3\)):
\begin{align}
    \bm d_1 &= a (0,1/\sqrt{3})\,,     \\
    \bm d_2 &= -a (1/2,1/2\sqrt{3})\,, \\
    \bm d_3 &= a (1/2,-1/2\sqrt{3})\,.
\end{align}
The tight-binding Hamiltonian consists in two terms, the hopping term with hopping energy \(\nu\), and the impurity term that couples electrons and holes with (classical) magnetic moments through an exchange constant \(J_I\) \cite{Uchoa-2011fk},
\begin{multline}
    \label{e:H}
    H(t)  = -\nu \sum_{<i,j>}\left( 
    \E^{-\I \phi_{ij}(t)} c_j^\dag \sigma_0 c_i +
    \E^{\I \phi_{ij}(t)} c_i^\dag \sigma_0 c_j\right) + \\
    J_I \sum_{i\in I} \bm n_i \cdot (c_i^\dag \bm \sigma c_i)\,,
\end{multline}
where \(c_i=(c_{i \uparrow}\; c_{i \downarrow})^T\) is the column annihilation operator of a particle of spin up (\(\uparrow\)) or down (\(\downarrow\)) at site \(i\). In order to preserve the translational symmetry, the external electric field \(\bm E\), is introduced through a time \(t\) dependent vector potential, \(\bm A = t \bm E = (0,-t E_y)\), giving the phase factor with \(\phi_{ij}(t) = -et \bm E \cdot \bm d_{{ij}}\), where \(e\) is the elementary charge, in the hopping term. In the impurity term,  \(\bm \sigma = (\sigma_x, \sigma_y, \sigma_z)\) stands for the Pauli matrices, and \(\sigma_0\) for the identity matrix; the sum spans over the set \(I\) of \(N_I\) randomly distributed sites, and \(\bm n_i\) is a normal vector pointing in the direction of the impurity magnetic moment. The number of impurities per site is denoted \(n_I=N_Ia^2/L^2\).

The Hamiltonian is suitably written in momentum space, such that the time dependent term is diagonal:
\begin{equation}
    \label{e:Ht}
    H(t) = \sum_{\bm k \in \mathit{BZ}} \psi^\dag_{\bm k} H_{\bm k}(t) \psi_{\bm k} + 
    \sum_{\bm k, \bm q \in \mathit{BZ}} \psi^\dag_{\bm q} V_{\bm q, \bm k} \psi_{\bm k}
\end{equation}
where \(\bm k\) is a wavenumber in the Brillouin zone \(\mathit{BZ}\) and 
\[
    \psi_{\bm k} = (\psi_{\bm k,A,\uparrow} \;
    \psi_{\bm k,B,\uparrow} \;
    \psi_{\bm k,A,\downarrow} \; 
    \psi_{\bm k,B,\downarrow})^T
\] 
is the annihilation operator of a particle having wavenumber \(\bm k\), belonging to the sublattice \((A,B)\), and of spin \(\sigma = \uparrow,\downarrow\). In momentum space, the hopping term of the Hamiltonian becomes,
\begin{equation}
    \label{e:Hk}
    H_{\bm k}(t) = \sigma_0\otimes \begin{pmatrix} 
        0 & h_{\bm k}(t) \\ 
h_{\bm k}^*(t) & 0
\end{pmatrix}
\end{equation}
where
\begin{equation}
    \label{e:hk}
    h_{\bm k}(t) = -\nu \sum_{a=1}^3 
    \E^{-\I (\hbar\bm k + et \bm E) \cdot \bm d_a}\,, \quad
    h_{\bm k}(0)=h_{\bm k}
\end{equation}
and the itinerant-fixed spin coupling term is given by the convolution
\begin{equation}
    \label{e:Vk}
    V_{\bm q, \bm k} =  J_I
    \sum_ {i \in I}\E^{-\I \bm q \cdot \bm x_i} 
    \bm n_i \cdot \bm \sigma \otimes \chi_i
    \E^{\I \bm k \cdot \bm x_i}
\end{equation}
where \(\chi_i=\mathrm{diag}(1,0)\) if \(i\in A\) and \(\chi_i=\mathrm{diag}(0,1)\) if \(i\in B\).

It is convenient to use \(\nu=3\,\mathrm{eV}\) and \(a=0.25\,\mathrm{nm}\) as the units of energy and length respectively; the unit of time is \(t_0=\hbar/\nu\approx 0.3\, \mathrm{fs}\), and the unit of electric field \(E_{0}=\nu/ea\approx 10^{10}\, \mathrm{V\,m^{-1}}\). The Fermi velocity is of the order \(v_F\sim \nu a/\hbar \approx 10^6\, \mathrm{ms^{-1}}\). In the following we use the system of units where \(\nu=a=\hbar=e=1\). Typical values of the model nondimensional parameters are taken as: \(J_I=0.1,\ldots, 1.5\), \(n_I=0.4\) \cite{Rappoport-2011vn}, and \(E_y=10^{-3}, \ldots, 10^{-2}\). 

The energy spectrum of the isolated clean system is given by the eigenvalues of \(H_{\bm k}(0)\) \cite{Dora-2010kx},
\begin{equation}
    \label{e:ek}
    E = \pm\epsilon_{\bm k}\,,\quad
    \epsilon_{\bm k} = \left| h_{\bm k} \right|\,,
\end{equation}
(contours of \(\epsilon_{\bm k}\) are represented in Fig.~\ref{f:1}). The corresponding eigenvectors are,
\begin{align*}
    \label{e:psii}
|\bm k,+,\uparrow\rangle &= \frac{1}{\sqrt{2}}\begin{pmatrix} 
\E^{\I  \phi_{\bm k}/2} \\ 
\E^{-\I \phi_{\bm k}/2} \\ 
0 \\ 
0 
\end{pmatrix},
& |\bm k,-,\uparrow\rangle &= \frac{1}{\sqrt{2}}\begin{pmatrix} 
-\E^{\I \phi_{\bm k}/2} \\ 
\E^{-\I \phi_{\bm k}/2} \\ 
0 \\ 
0 
\end{pmatrix}, \\
|\bm k,+,\downarrow\rangle &= \frac{1}{\sqrt{2}}\begin{pmatrix} 
0 \\ 
0 \\ 
\E^{\I  \phi_{\bm k}/2} \\ 
\E^{-\I \phi_{\bm k}/2} 
\end{pmatrix},
& |\bm k,-,\downarrow\rangle &= \frac{1}{\sqrt{2}}\begin{pmatrix} 
0 \\ 
0 \\ 
-\E^{\I \phi_{\bm k}/2} \\ 
\E^{-\I \phi_{\bm k}/2}
\end{pmatrix},
\end{align*}
where \(\tan\phi_{\bm k} = \mathrm{Im}\, h_{\bm k}/ \mathrm{Re}\, h_{\bm k}\), and the signs \(\pm\) correspond to positive (electrons) or negative (holes) energy states.

The time evolution of the system is computed using a splitting method in momentum space:
\begin{equation}
    \label{e:tsa}
    \Psi_{\bm k}(t+\Delta t) = U_{\bm k}\big(\tfrac{1}{2}\Delta t\big)
    T_{\bm k}(t+\Delta t)
    U_{\bm k}\big(\tfrac{1}{2}\Delta t\big) \Psi_{\bm k}(t)\,,
\end{equation}
accurate to second order in the time step \(\Delta t\), where
\[
    T_{\bm k}(t+\Delta t) = \exp\big\{ -\tfrac{\I}{2}  \Delta t
    [H_{\bm k}(t+\Delta t) + H_{\bm k}(t)] \big\}\,,
\]
and
\[
    U_{\bm k}\big(\tfrac{1}{2}\Delta t \big) = F_{\bm k,i}^{-1} \circ 
    \exp\big[- \tfrac{\I}{2}\Delta t V_i \big] \circ F_{i,\bm k}\,,
\]
with \(F_{i,\bm k}\) denoting the Fourier transform, and \(V_i = J_s \bm n_i \cdot \bm \sigma \otimes \chi_i\) the impurity potential energy; the wavefunction is obtained from \(\Psi(\bm x_i,t) = F_{i,\bm k} \circ \Psi_{\bm k}(t)\). The mesh of vectors \(\bm k\) is defined in the rectangle of Fig.~\ref{f:1}, having twice the area of the first Brillouin zone.

%
% FIG 3
\begin{figure}
\centering
\includegraphics[width=0.48\textwidth]{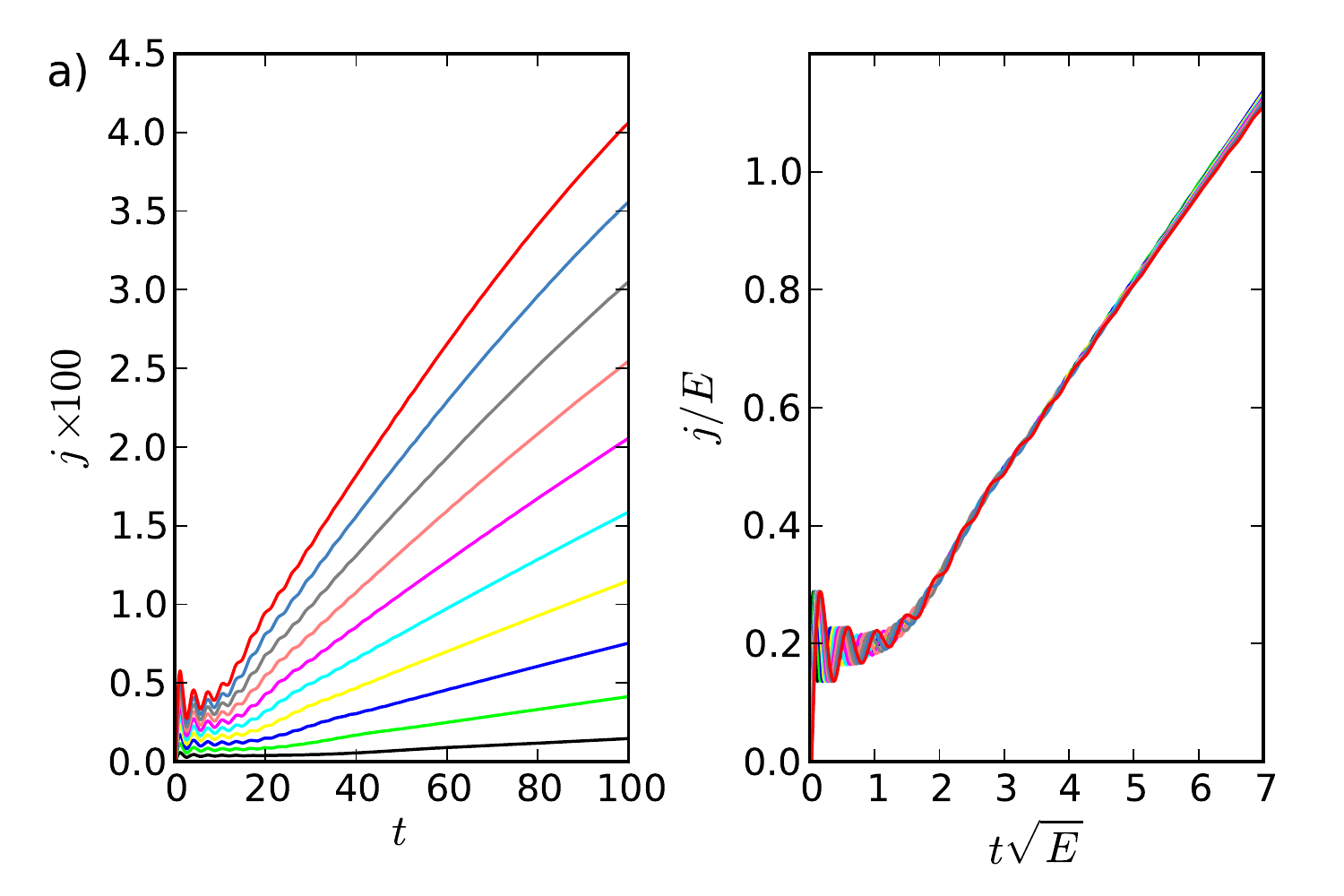}\\
\includegraphics[width=0.48\textwidth]{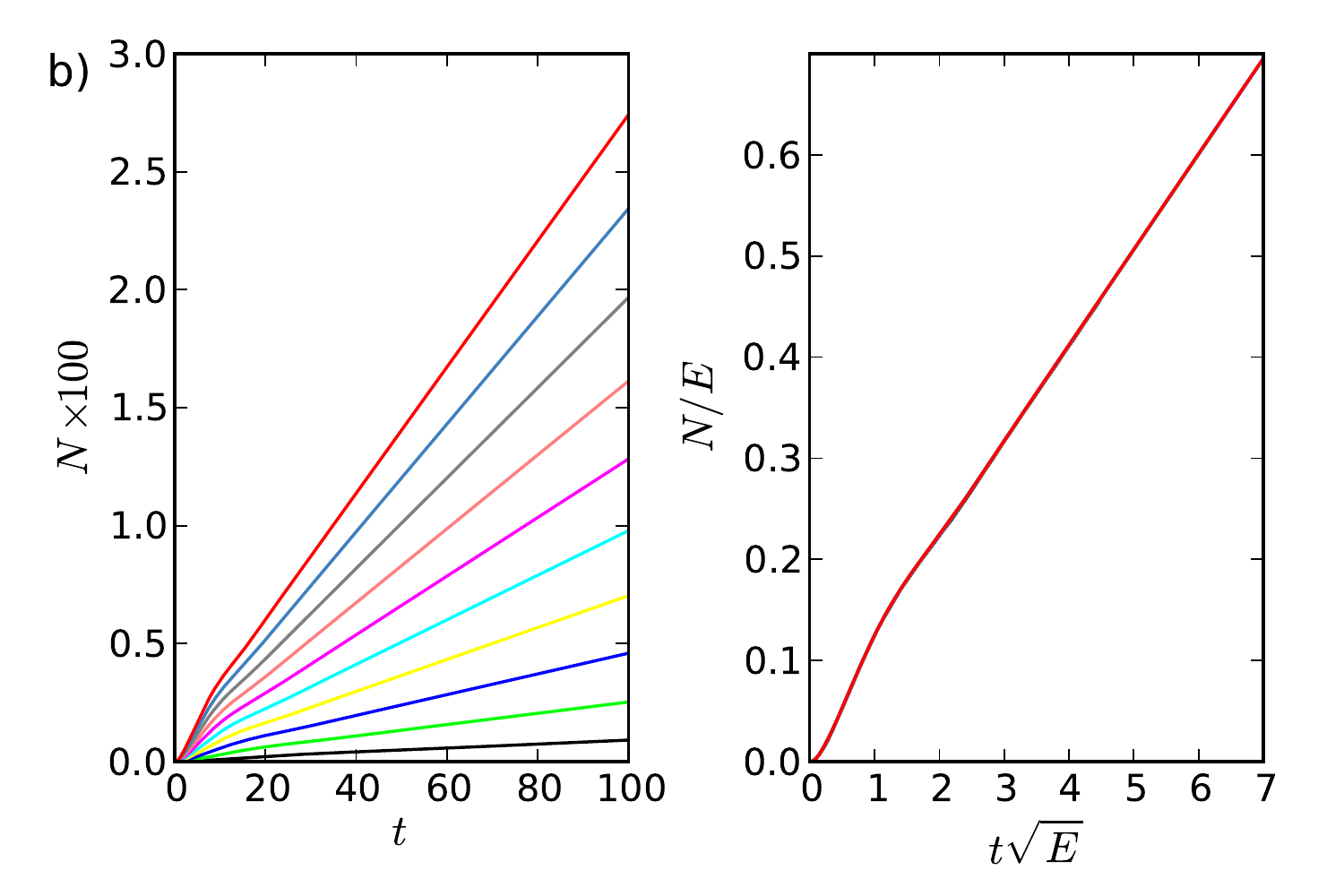}
\caption{\label{f:3} (color online). Current (a) and pair creation rate (b), for a clean graphene sheet. When the data is scaled with appropriated powers of the electric field, it collapses to a single curve (as shown on the right panels). Electric field \(E=0.002,0.004,\ldots,0.02\, E_0\) (\(10\) values in steps of $0.002\, E_0$, from black to red).}
\end{figure}

To illustrate the behavior of the system in the simplest case, we show in Fig.~\ref{f:2}, snapshots of the hole and electron probability densities \(|\Psi(\bm x_i, t)|^2\) for the clean system, at different times. A logarithm scale is used to enhance the small values of the wave function. Initially a hole is put at the center of the lattice, in a state with wave function \(\Psi(\bm x_0, 0) = \langle \bm x_0, 0|\bm k, -, \uparrow \rangle\). The initial electron wave function is zero, but as shown in the left panel of Fig.~\ref{f:2}, it increases with time. The maximum of the probability density tends to drift in the direction of \(E_y\) for the holes, and in the opposite direction for the electrons. While the electron density increases in the positive \(y\)-direction, the hole density develops simultaneously an asymmetry, with a larger concentration in the \(-y\)-direction. The growth of the electron density is related to the creation of electron-holes pairs. Indeed, under the effect of the strong electric field, electron-hole pairs are produced through the Schwinger mechanism \cite{Schwinger-1951fk,Dora-2010kx,Rosenstein-2010uq}, leading to a nonlinear response regime. Remark that at times \(t\approx 60\, t_0\) the wave function reach the borders of the system, given an order of magnitude for the threshold of finite size effects (that depend on the strength of the electric field); in the following we show the evolution of the physical quantities up to times \(t=100\,t_0\).

In order to characterize the transport in this regime or in the presence of impurities, we monitor the pair creation rate \cite{Kao-2010fk},
\begin{equation}
    \label{e:N}
    N(t) = \sum_{\bm k \in \mathit{BZ}} 
    \left\langle \left| \langle -| 
    \psi_{\bm k}^\dag(t)\, \psi_{\bm k} |+\rangle \right|^2 
    \right \rangle\,,
\end{equation}
where \( |+\rangle = |\bm k,+,\uparrow\rangle\), and \(|-\rangle = |\bm k,-,\uparrow\rangle\), as well as the mean current density (averaged over the area \(L^2\)),
\begin{equation}
    \label{e:j}
    j(t) = \langle j_y \rangle(t) = -\frac{1}{L^2} \sum_{\bm k \in \mathit{BZ}} 
    \left\langle  \langle 0| \psi^\dag_{\bm k}(t)\,
    \frac{\partial H_k}{\partial A_y}\, 
    \psi_{\bm k}(t) |0 \rangle \right\rangle\,,
\end{equation}
where the external brackets \(\langle\cdots\rangle\) are for the disorder averaging, and \(|0\rangle\) is the initial state, usually taken to be \(|0\rangle = |\bm k,-,\uparrow\rangle\) (a spin-up hole centered at the origin). The current density and the corresponding pair creation rate, in the clean case, are represented in Fig.~\ref{f:3}, for different values of the electric field. After an initial transient, in which the current oscillates around a constant and whose duration is shorter with increasing fields, the current grows almost linearly in time. The constant characterizing the initial regime \(j(t)/E\), corresponds to the conductivity,
\begin{equation}
  \label{e:sigma0}
  \sigma_0 = \frac{4}{\pi} \frac{e^2}{h}=\frac{2}{\pi^2}\,,
\end{equation}
obtained from the linear response theory for static fields \cite{Peres-2010uq}. A straightforward calculation, using for instance the analogy of the low energy Dirac system with the Hamiltonian of the Landau-Zener tunneling \cite{LandauQM}, leads to the scaling \(t\rightarrow \sqrt{E} t\) and \(j \rightarrow E (\sqrt{E} t)\); the pair creation rate behaves similarly. Explicitly one obtains \cite{Dora-2010kx},
\begin{equation}
  \label{e:slope}
  j(t)/E = 2e v_F N(t)/E =  \sigma_0 \sqrt{v_F E} t\,,
\end{equation}
where, in our units, the Fermi velocity near the Dirac point is \(v_F=\sqrt{3}/2\). These scalings are confirmed numerically, as shown in the plots of Fig.~\ref{f:3} (right column); in particular, the slope predicted by Eq.~(\ref{e:slope}), \(0.189\) is only slightly larger than the numerical result, about \(0.16\). The difference may be attributed to a renormalization of the continuous, low energy formula (\ref{e:slope}), due to the lattice and its intrinsic length scale \(a\) and finite energy band width. In addition, the corresponding slope of the pair creation rate, found to be about \(0.1\), is in perfect agreement with the relation \(j(t)/N(t)=2ev_F=\sqrt{3} \approx 1.6\). Therefore, the linear and nonlinear regimes are both characterized by the \emph{same} prefactor, the static conductivity \(\sigma_0\). In addition, as we demonstrate in the following section, the behavior observed in the clean limit extends smoothly to the (parametric) disorder regime, as predicted by the linear response theory: the static conductivity is insensitive to weak disorder. 

%
% FIG 4
\begin{figure}
\centering
\includegraphics[width=0.48\textwidth]{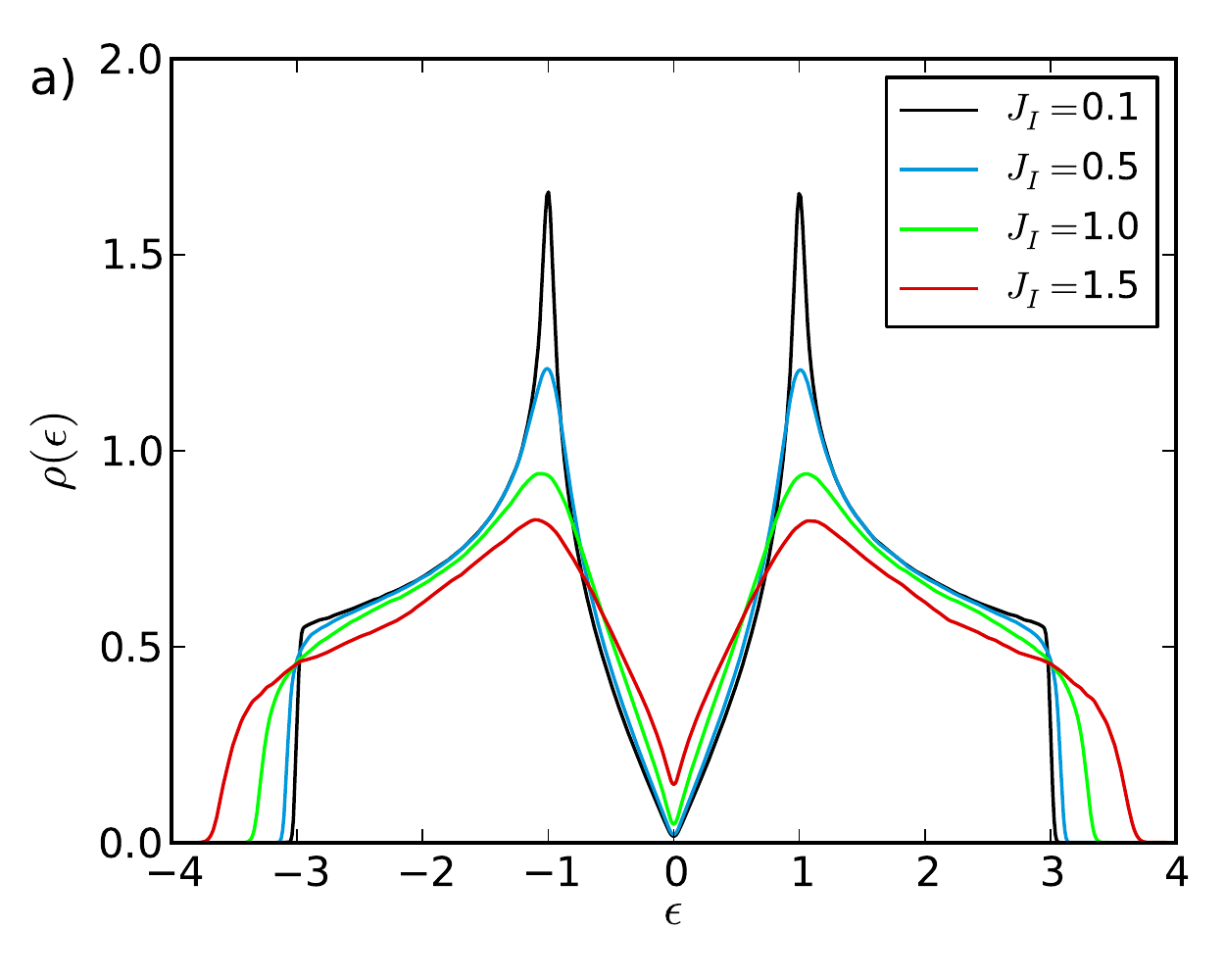}\\
\includegraphics[width=0.48\textwidth]{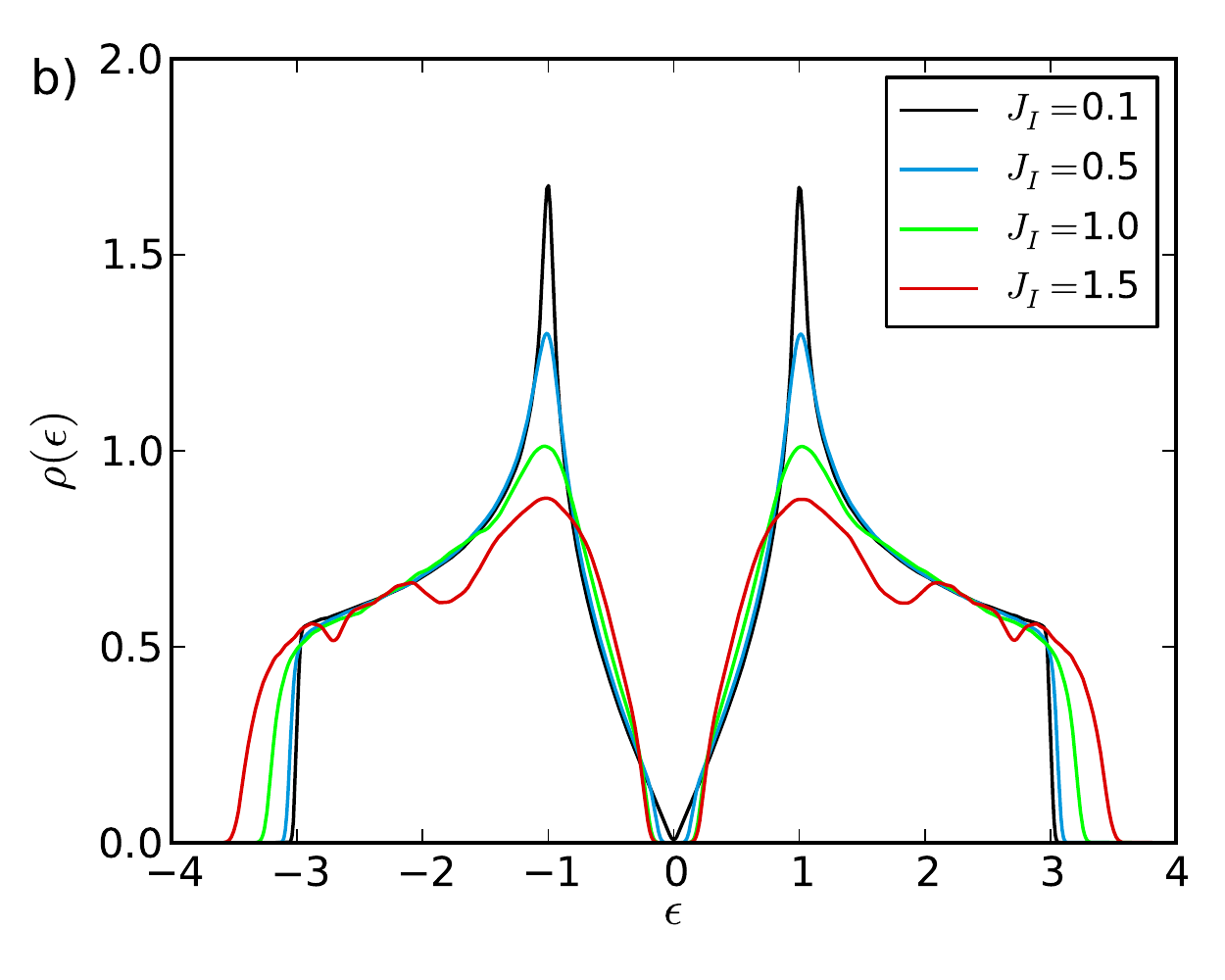}
\caption{\label{f:4} (color online). Effect of disorder on the density of states. (a) paramagnetic case; (b) magnetic case. Disorder \(J_I=0.1,0.5,1.0,1.5\,\nu\).}
\end{figure}

It is worth mentioning that the clean static regime is singular within the framework of the linear response approximation, in the sense that the value of the conductivity depends on the specific way the zero frequency \(\omega\) and the disorder strength limits are taken (as already noted in the seminal paper of Ref.~\onlinecite{Ludwig-1994fk}). Indeed, in the low frequency limit instead of the static value (\ref{e:sigma0}), one obtains,
\[
  \bar{\sigma}=\lim_{\omega \to 0} \sigma(\omega)=(\pi/2) (e^2/h)=1/4 \ne \sigma_0\,.
\] 
This value of the conductivity was found elsewhere for the initial linear regime, using an approximation valid for finite momentum \(p \gg eEt\),\cite{Lewkowicz-2009kx} or more generally, for the whole linear and nonlinear regimes, using a truncated series representation of the solution of the Dirac equation, computed using small and large momentum cut-offs (see Ref.~\onlinecite{Kao-2010fk}).%
\footnote{Note however, that the nonlinear evolution found in Ref.~\onlinecite{Kao-2010fk} is identical to the one shown in Fig.~\ref{f:3}. In fact, the fitting formula (\ref{e:slope}) is rigourosly equivalent to Eq.~(72) of the referred paper (where the authors denote \(\sigma_2\lambda=\sigma_0\sqrt{v_F}=(2/\pi^2)(\sqrt{3}/2)^{1/2}\) with \(\sigma_2=\bar{\sigma}\)).} %
In the present model we use a spectral integration method that allows to exactly compute the differential operators on the lattice. The full account of the lattice effects regularize the dynamics, leading naturally to the conductivity \(\sigma_0\), in a strictly constant electric field, \(\omega=0\).

%
% FIG 5
\begin{figure*}
\centering
\includegraphics[width=0.16\textwidth]{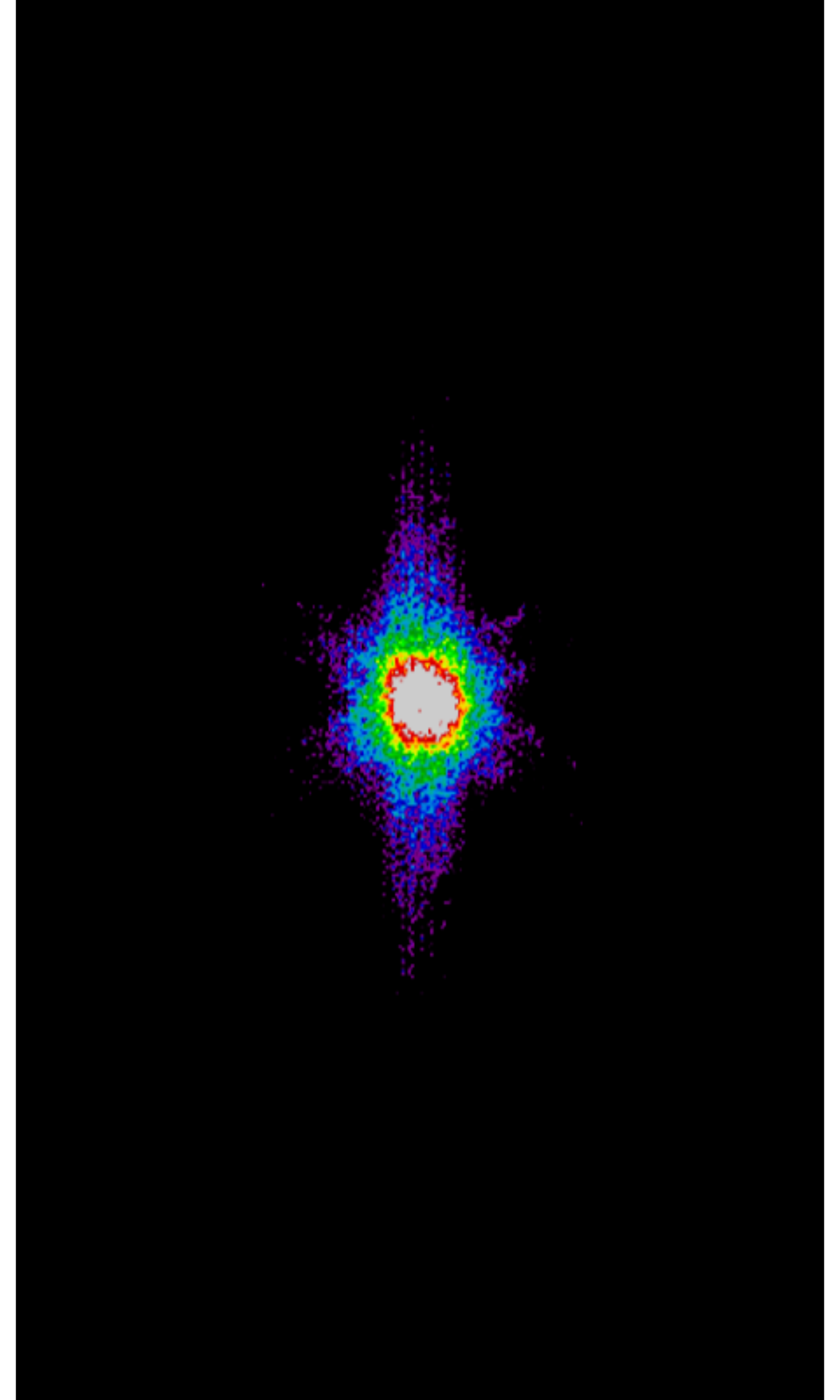}%
\includegraphics[width=0.16\textwidth]{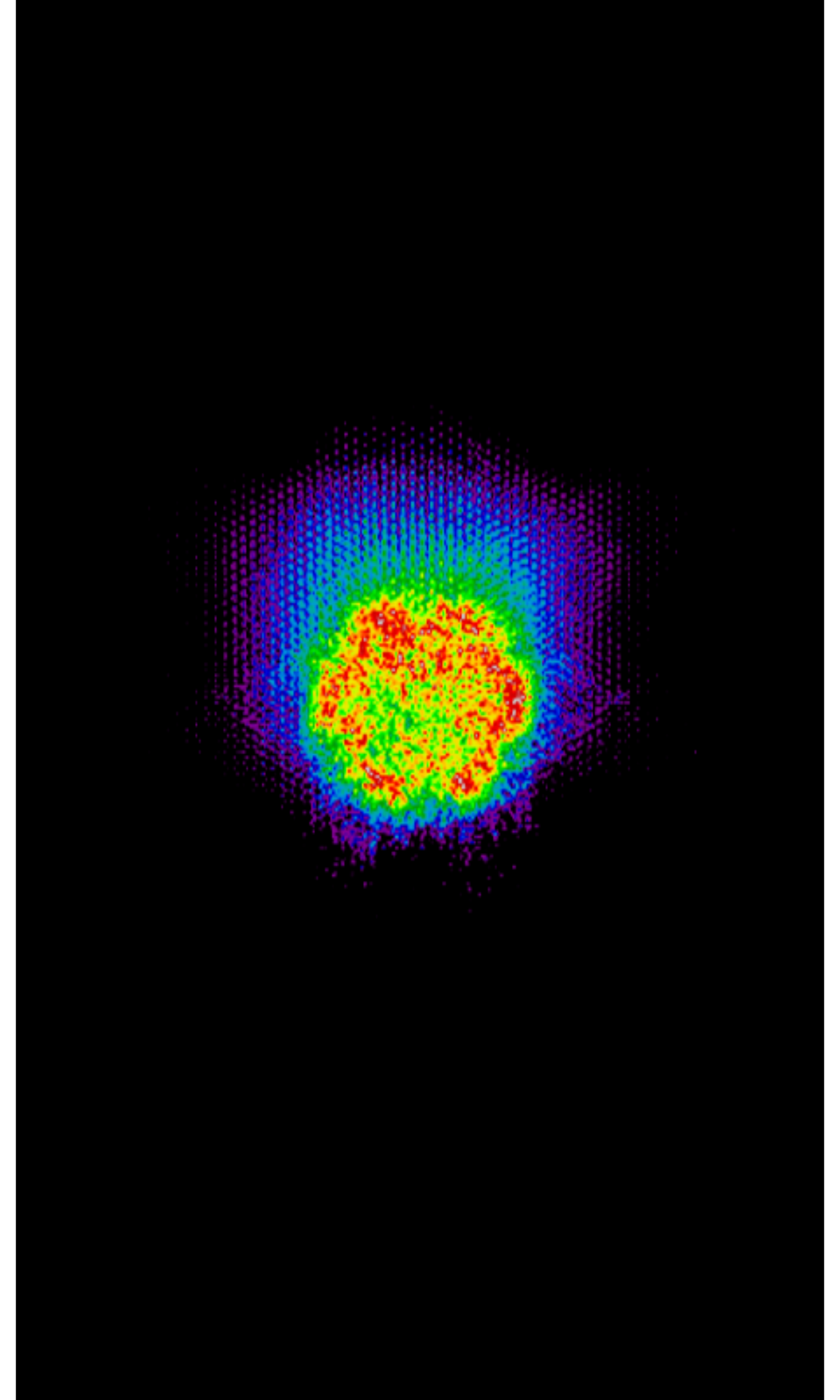}%
\includegraphics[width=0.16\textwidth]{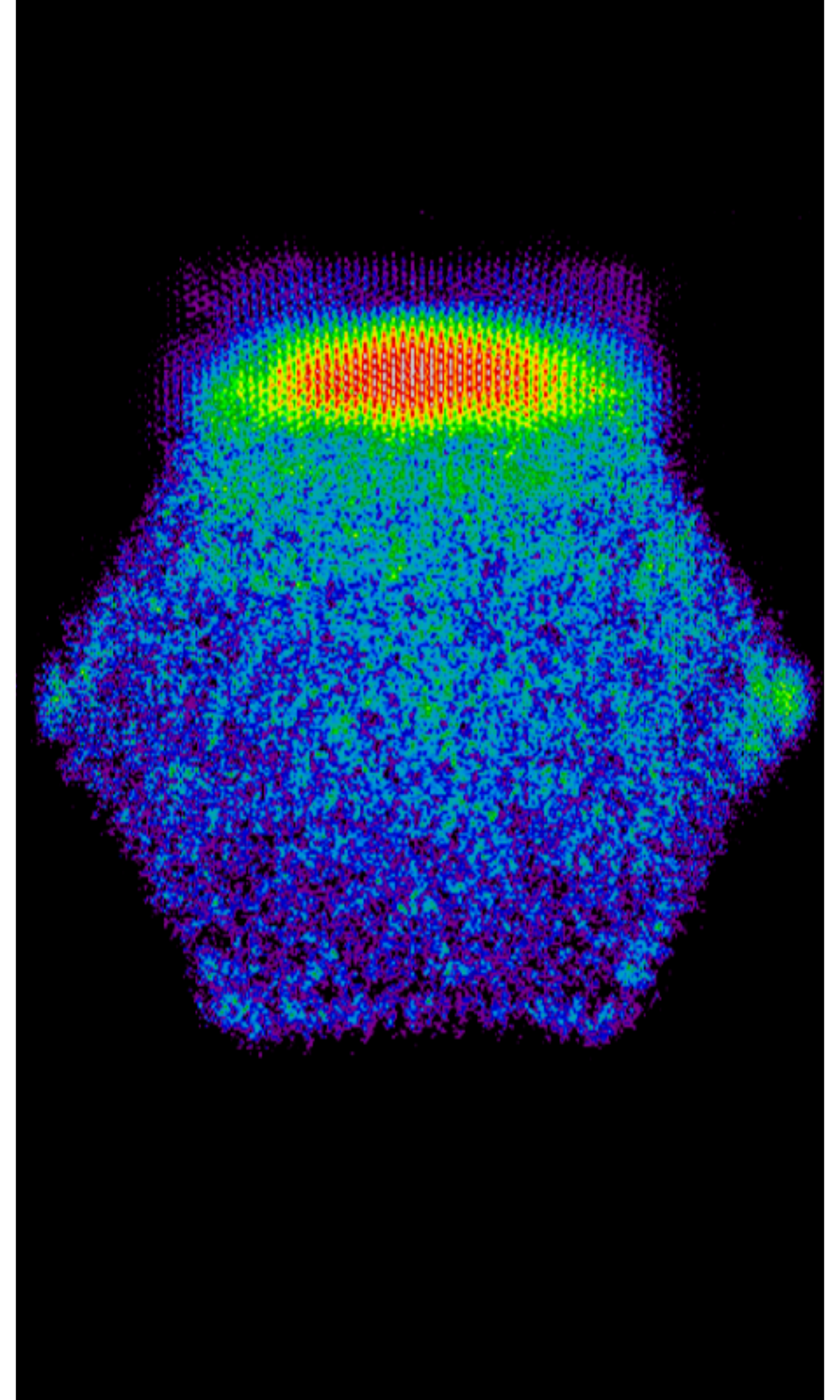}\hfill%
\includegraphics[width=0.16\textwidth]{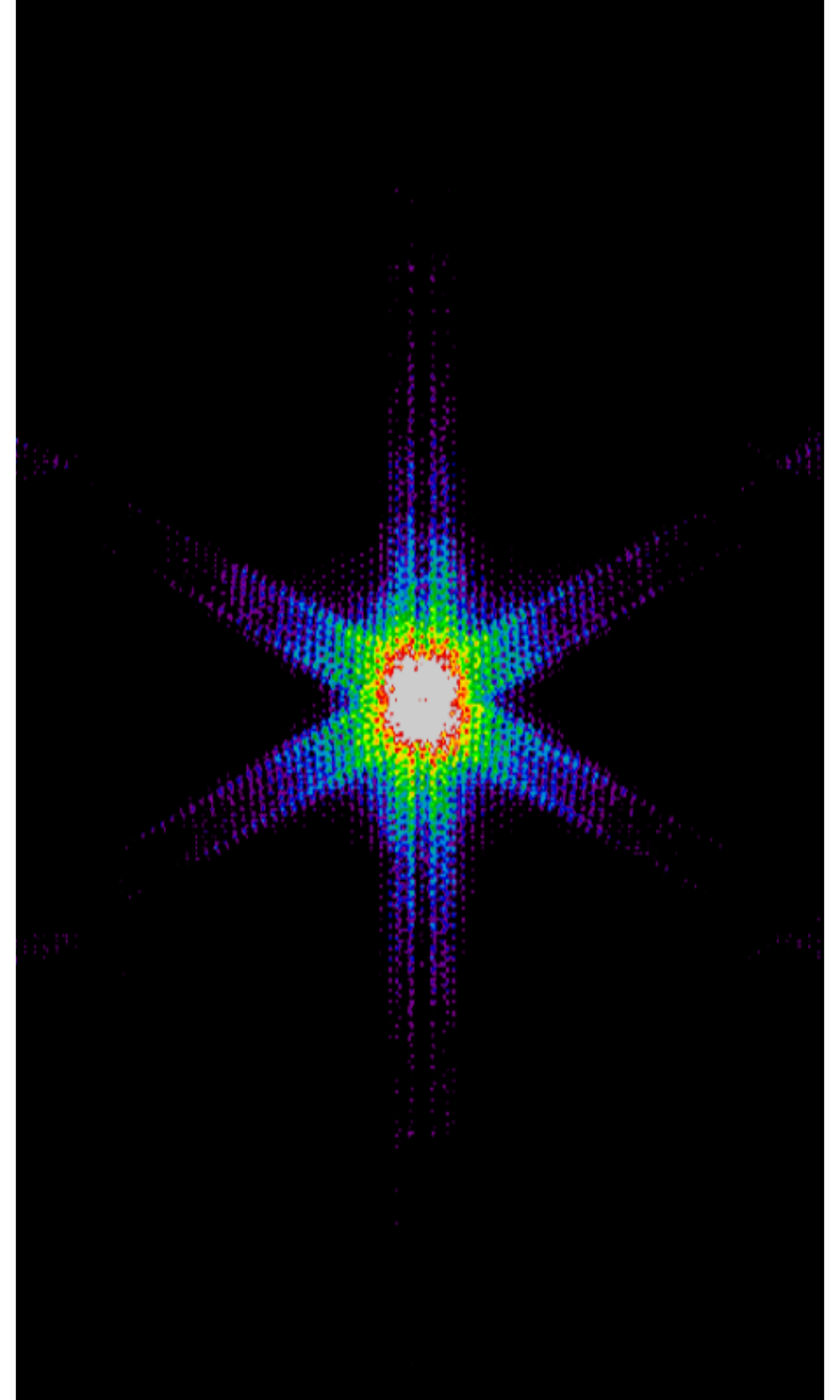}%
\includegraphics[width=0.16\textwidth]{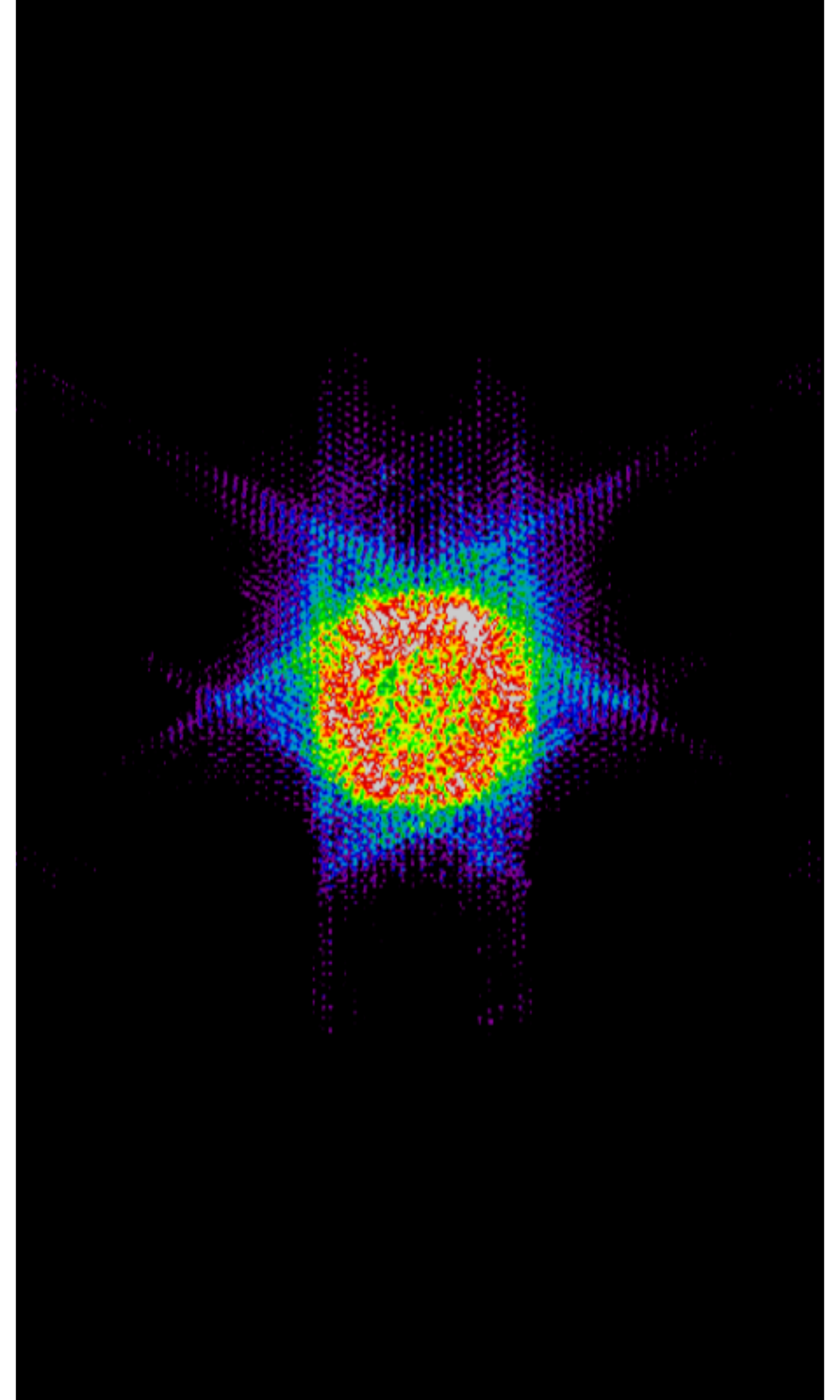}%
\includegraphics[width=0.16\textwidth]{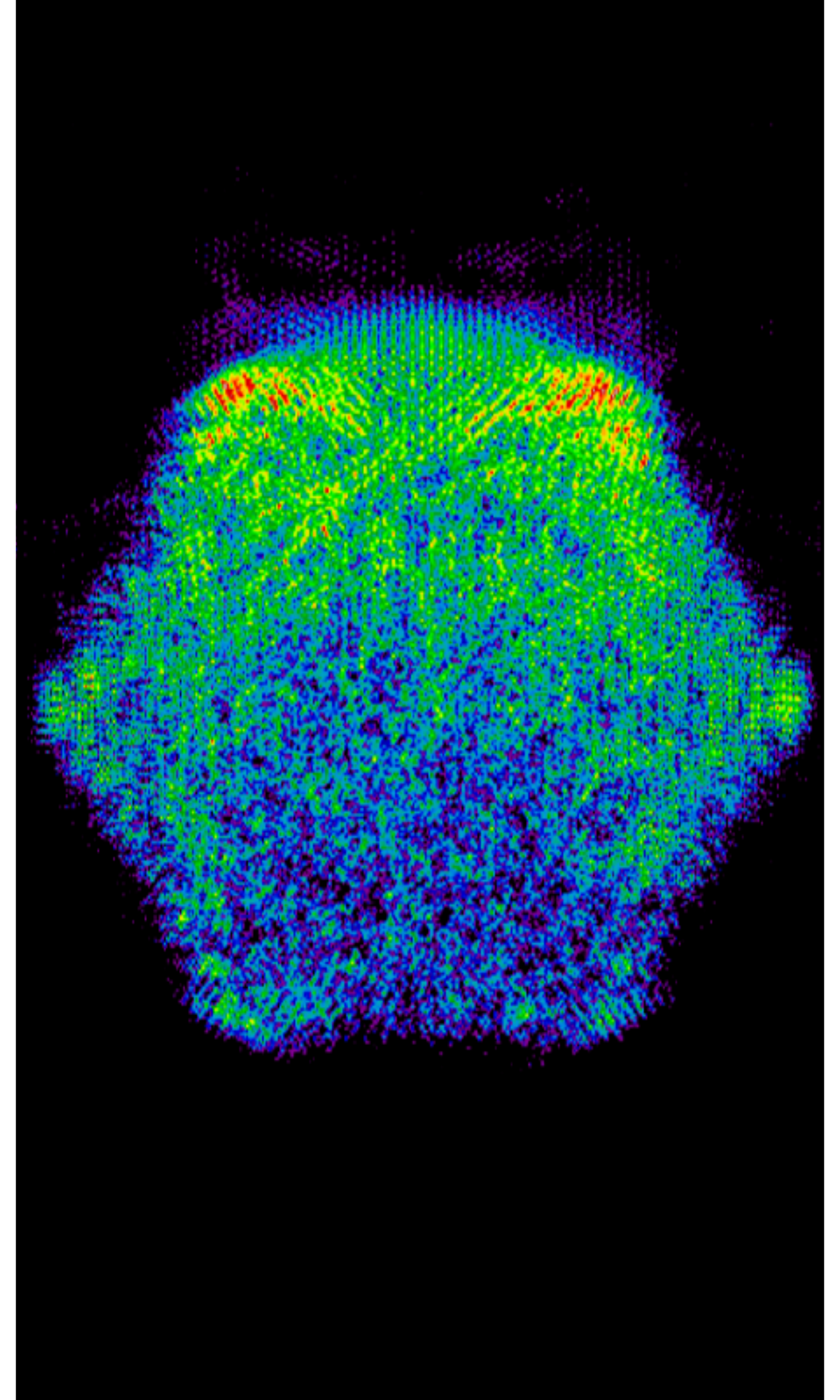}\\
\includegraphics[width=0.16\textwidth]{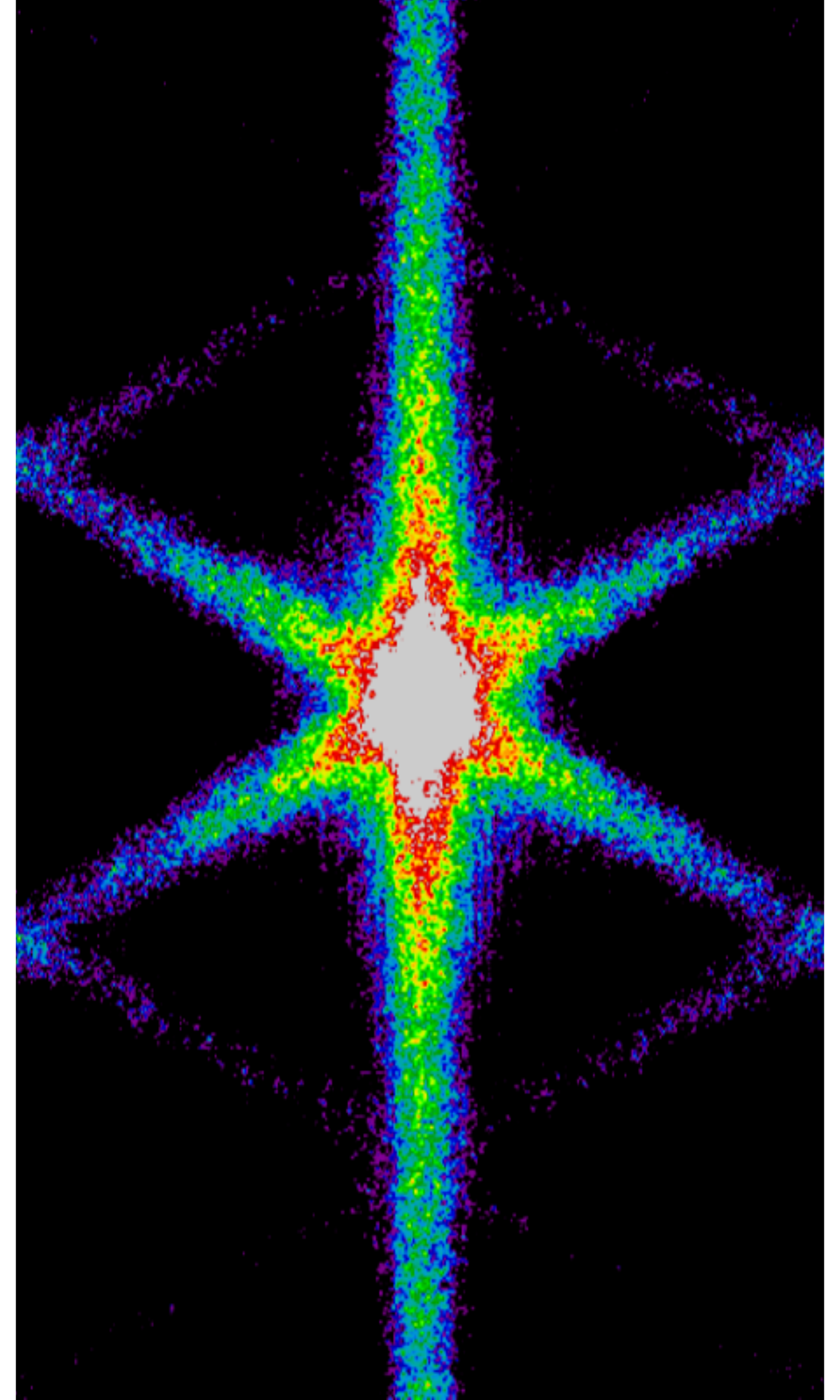}%
\includegraphics[width=0.16\textwidth]{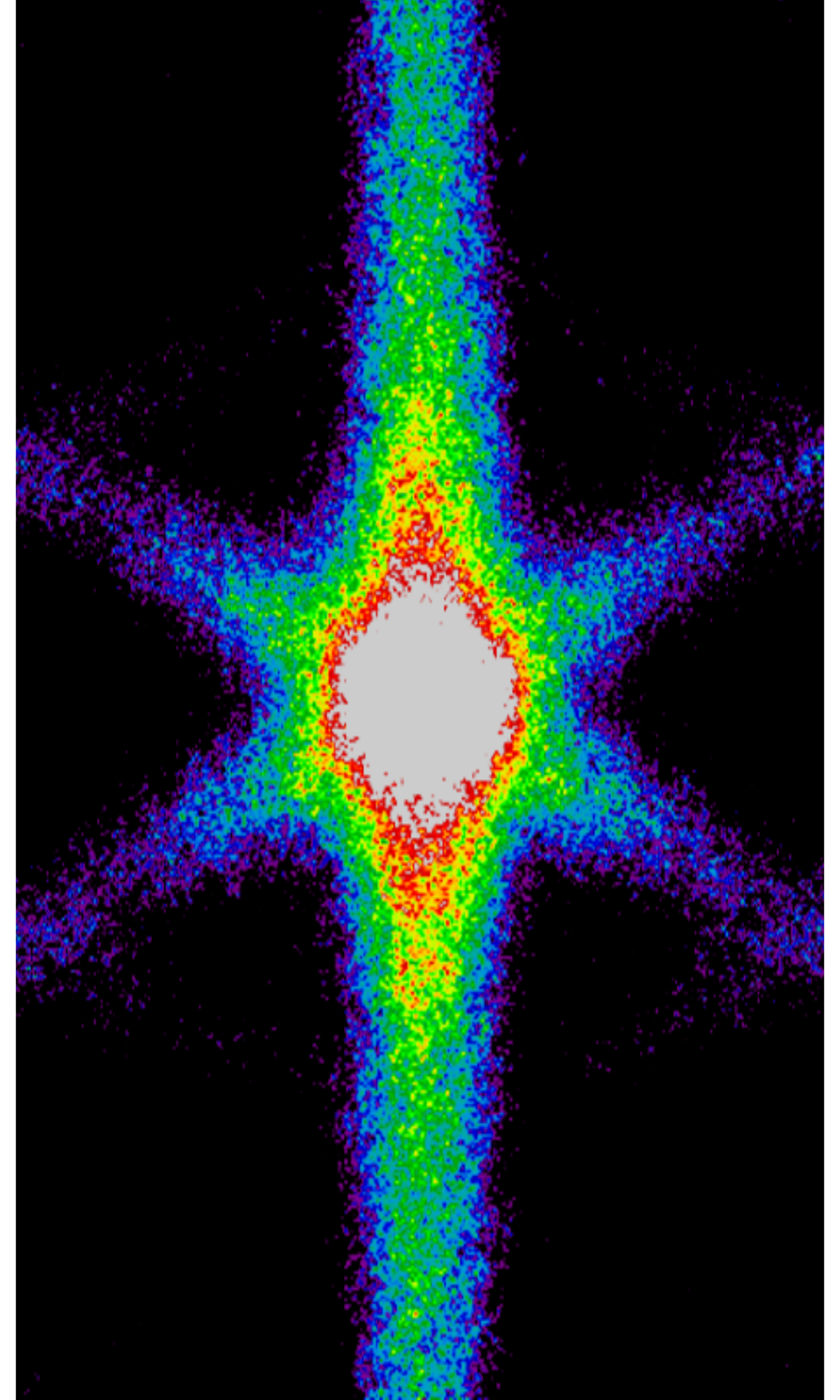}%
\includegraphics[width=0.16\textwidth]{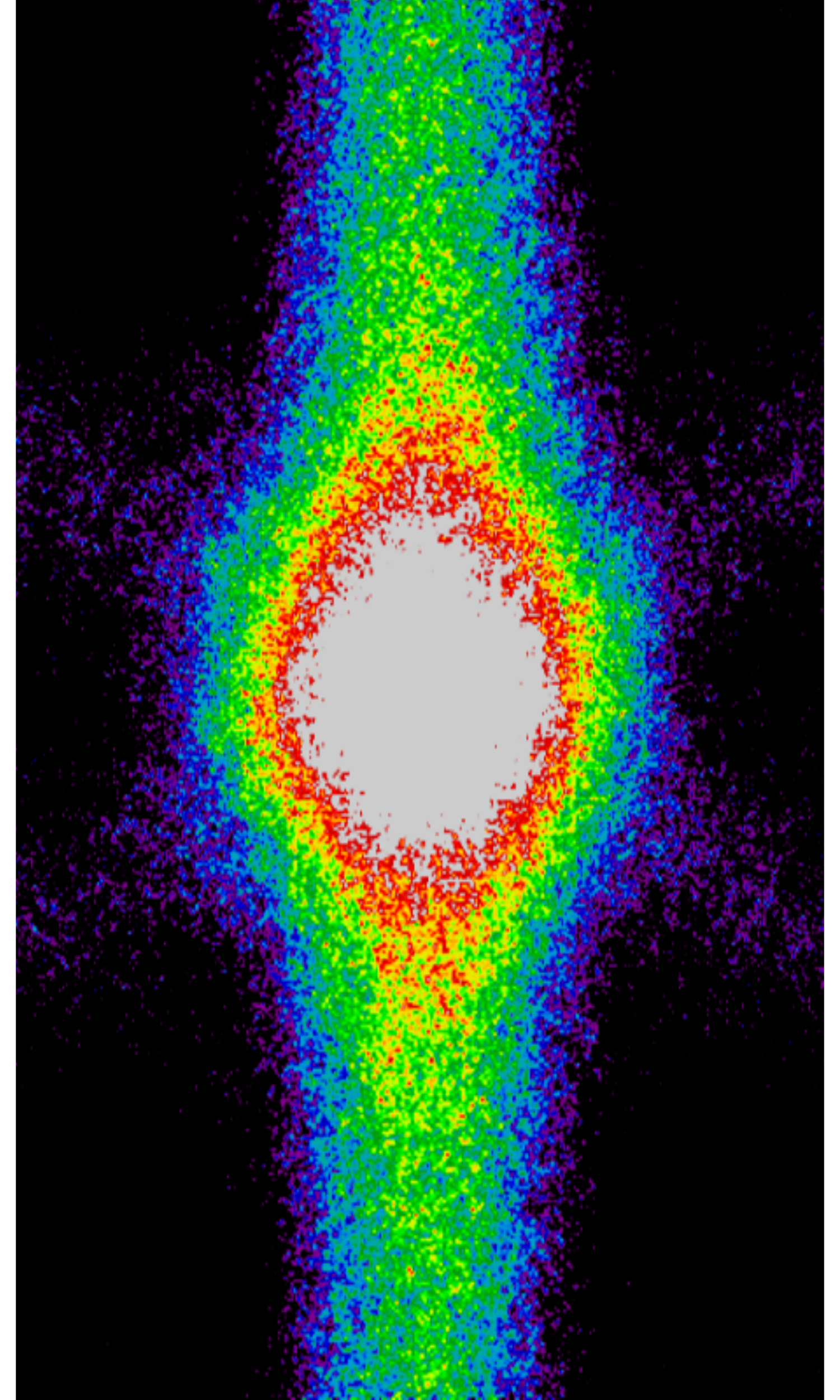}\hfill%
\includegraphics[width=0.16\textwidth]{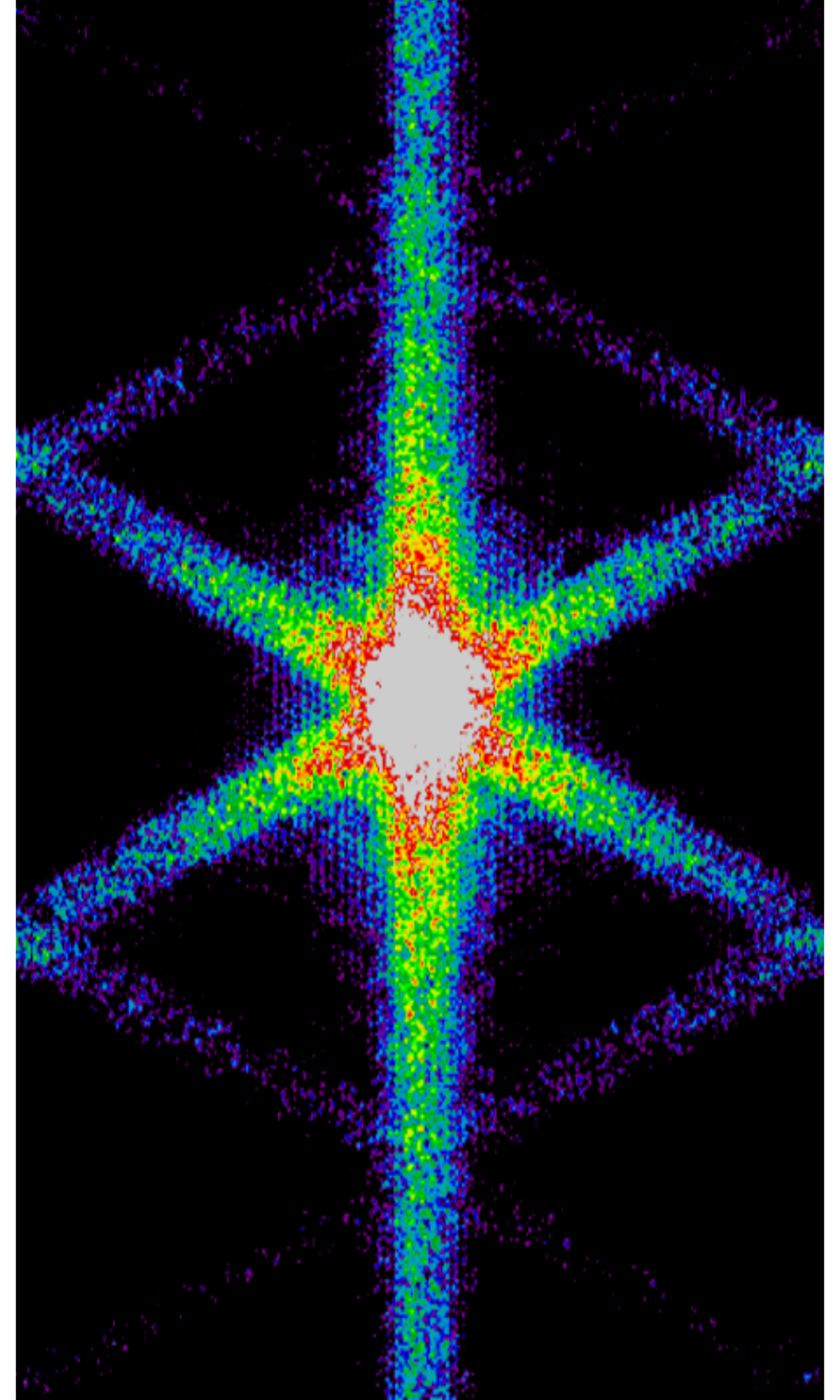}%
\includegraphics[width=0.16\textwidth]{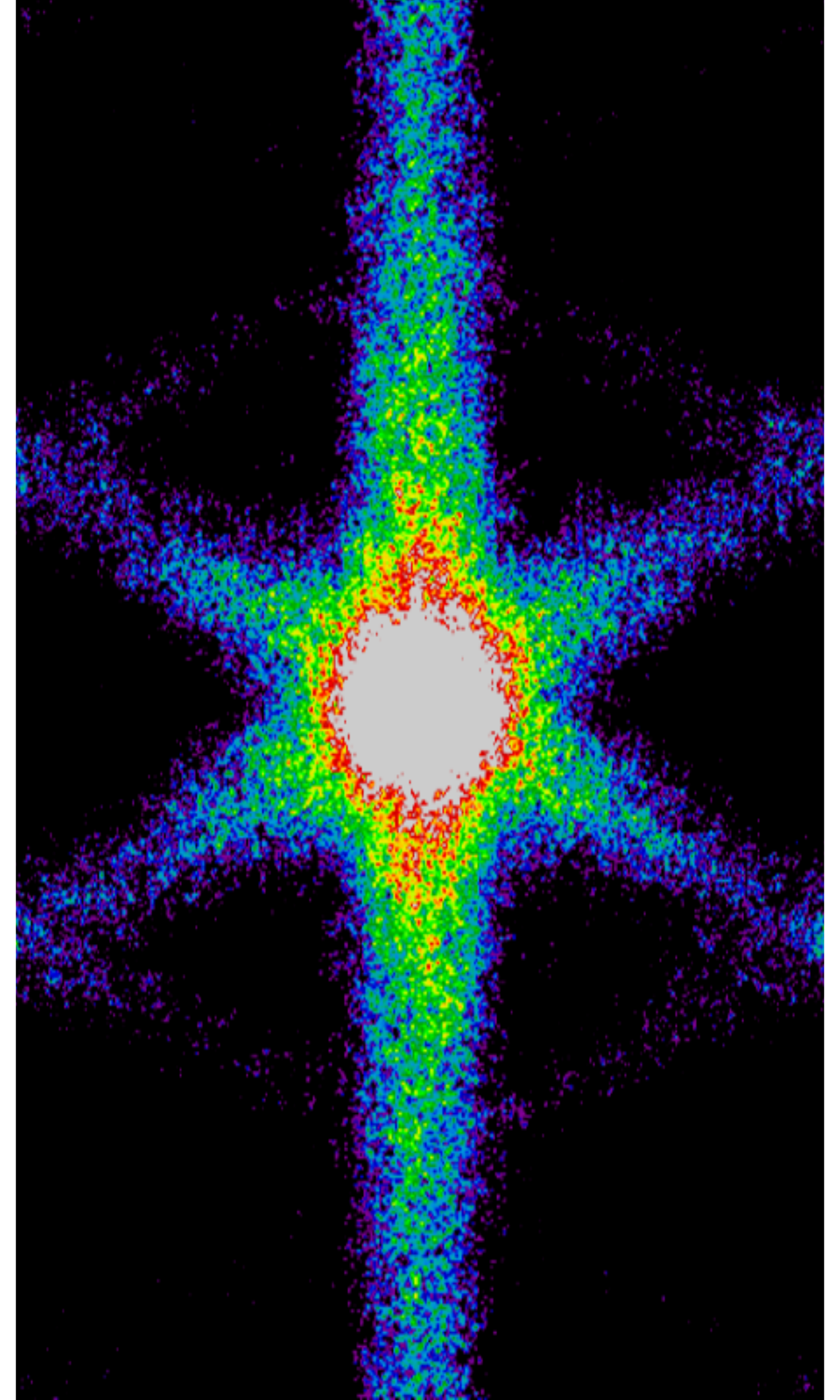}%
\includegraphics[width=0.16\textwidth]{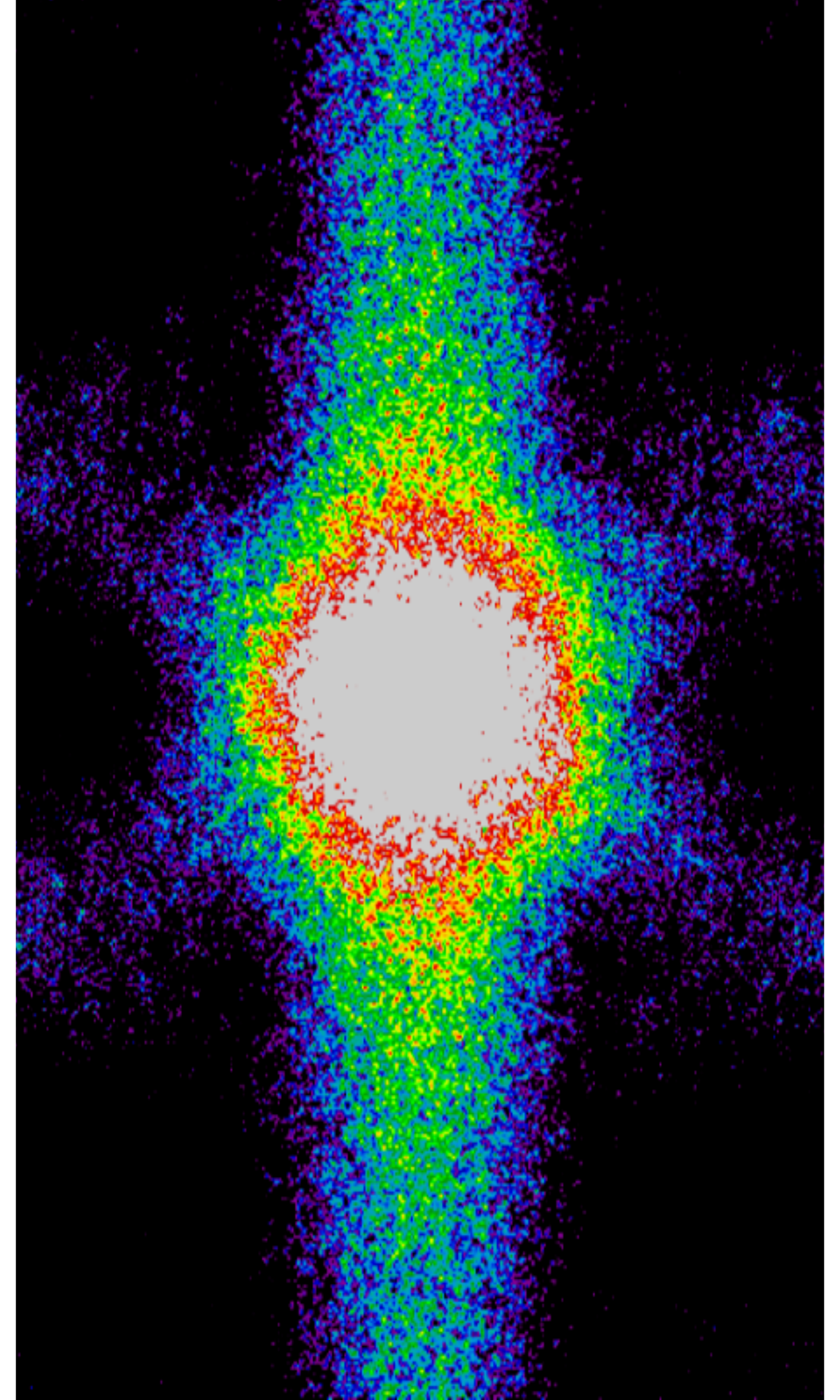}
\caption{\label{f:5} (color online). Electron density as a function of time for different disorder strengths (top \(J_I=0.1\, \nu\), bottom \(J_I=1.0\, \nu\)) and configurations (left paramagnetic, right magnetic). Electric field \(E=0.02\, E_0\), times \(t=4,16,60\, t_0\)), lattice size \((2 \times 256\, a)^2\).}
\end{figure*}

\section{Current and pair creation in the disordered system}
\label{s:current}

In the following we consider two types of magnetic disorder, one with the orientation of the magnetic moments \(\bm n_i\), uniformly distributed on the sphere, and the other with \(\bm n_i=(0,0\pm 1)\) for sites in the two sublattices A and B, respectively. The effect of impurities on the electronic bands will depend on these two types magnetic order: randomly oriented moments (paramagnetic case), will contribute to populate the energy levels around the Fermi energy; magnetic moments following an antiferromagnetic order with different spin orientation on the two sublattices (magnetic case), will break the time reversal symmetry and open a gap. A quantitative measure of these effects can be obtained from the density of states, 
\begin{equation}
    \label{e:dos}
    \rho(\epsilon) = \sum_n \left\langle \delta(\epsilon-\epsilon_n)
        \right\rangle\,,
\end{equation}
where \(\epsilon_n\) are the eigenstates of the disordered Hamiltonian. We compute the density of states Eq.~(\ref{e:dos}), using the Chebychev method \cite{Weisse-2006fk}. In Fig.~\ref{f:4} we show the density of states for increasing disorder strength (given by the values of the exchange constant), in both paramagnetic and magnetic cases. For increasing paramagnetic disorder, the energy band width extends and a finite density of states near \(\epsilon=0\) develops. For increasing magnetic disorder, the behavior near \(\epsilon=0\) change drastically: a gap whose width is proportional to the disorder strength, is created \cite{Daghofer-2010fk,Rappoport-2011vn}. One may anticipate that the type of disorder will influence the transport properties differently, according to the modification they may induce on the system symmetries; n particular, the magnetic order can change qualitatively the response of the system to the applied electric field, because of the breaking of the underlying time reversal symmetry.

Figure~\ref{f:5} presents the time evolution of the electron density for the paramagnetic case (left) and magnetic case (right), for two values of the disorder strength, weak (top) and strong (bottom). They can be compared with the clean case of Fig.~\ref{f:2} (left). At weak disorder, the electron spreads, as in the clean case, almost ballistically (top panels). Increasing the disorder strength results in a change of regime, towards a diffusive transport regime (bottom panels). It is also worth noting, that in the initial stage of the system evolution, the electron density rapidly increases, as compared with the clean case, suggesting an enhanced rate of pair production in the presence of impurities. The main effect of disorder is in the rapid and reinforced spreading of the probability density, due to the scattering off impurities. As a result the current must decrease, as part of the electron density drags behind the drifting maximum. The comparison of the two kinds of disorder reveals that in the ferromagnetic case the asymmetry of the distribution is smaller than in the paramagnetic case, and that for strong disorder it tends to become almost isotropic signaling a possible effect of localization.

%
% FIG 6
\begin{figure}
\centering
\includegraphics[width=0.48\textwidth]{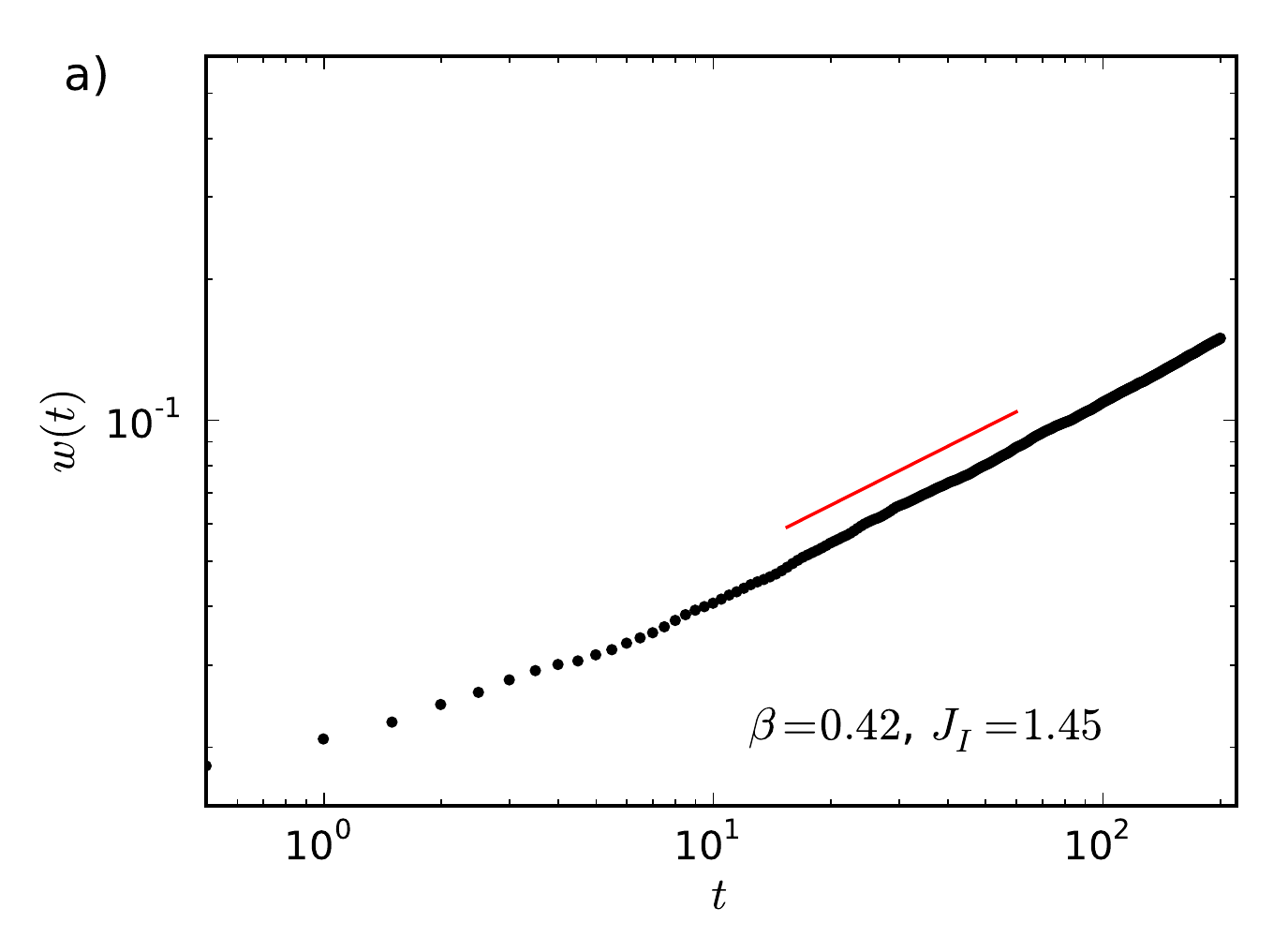}\\
\includegraphics[width=0.48\textwidth]{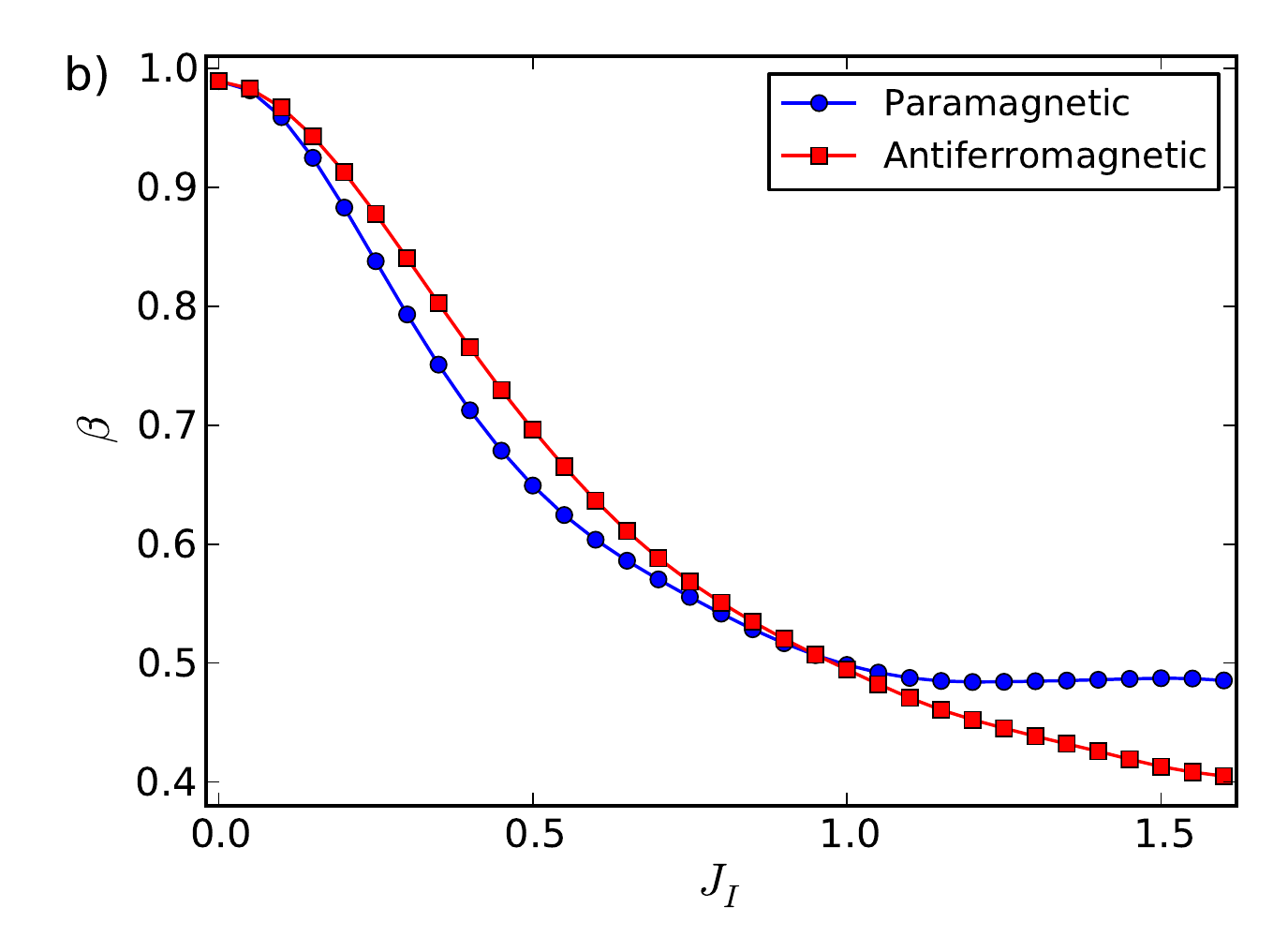}
\caption{\label{f:6} (color online). Width of a wave packet, propagating in a disordered environment (zero electric field). (a) Example of \(w(t)\) in the magnetic case, with a subdiffusive exponent \(\beta=0.45\) (the width is measured in unites of the lattice constant \(a\)). (b) Power law exponent as a function of the disorder strength.}
\end{figure}

The change between the ballistic and diffusive regimes is also supported by the measure of the wave function width \(w(t)\), 
\begin{equation}
    \label{e:w}
    w(t)^2 = \int_{L^2}\! d\bm x \, |\bm x|^2 |\Psi(\bm x,t)|^2
\end{equation}
represented in Fig.~\ref{f:6}. The wave packet evolves following a power law \(w(t)\sim t^\beta\), with characteristic exponent \(\beta\). The exponent that in the clean case has the ballistic value \(\beta=1\), tends in the paramagnetic case, to its diffusion value \(\beta=0.5\); for strong disorder, in the magnetic case, it shows a transition from superdiffusive to subdiffusive behavior. The transition between these two regimes coincides with the crossing of the curves in Fig.~\ref{f:6}b, at about \(J_I\approx 1\,\nu\).

%
% FIG 7
\begin{figure*}
\centering
\includegraphics[width=0.35\textwidth]{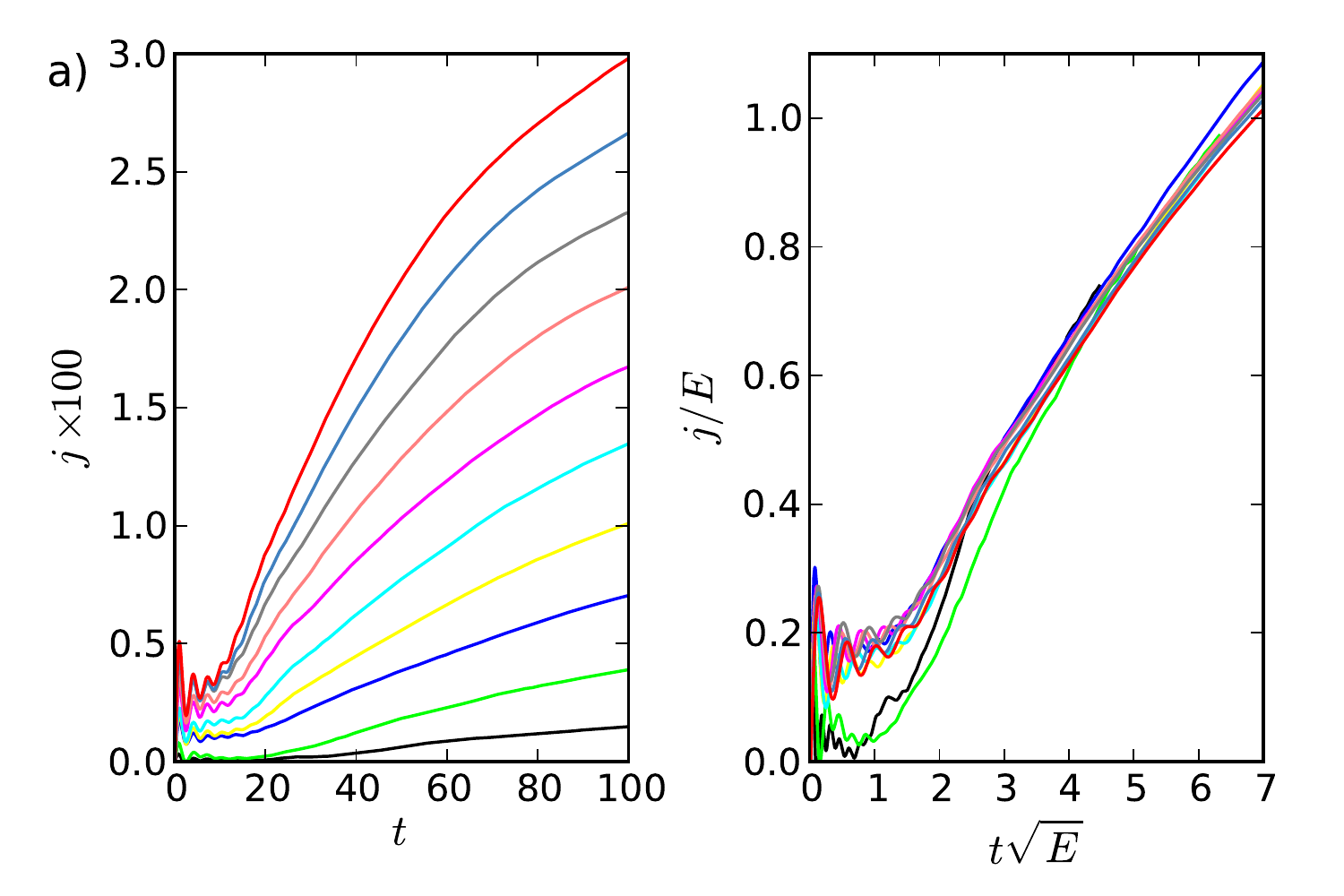}%
\includegraphics[width=0.35\textwidth]{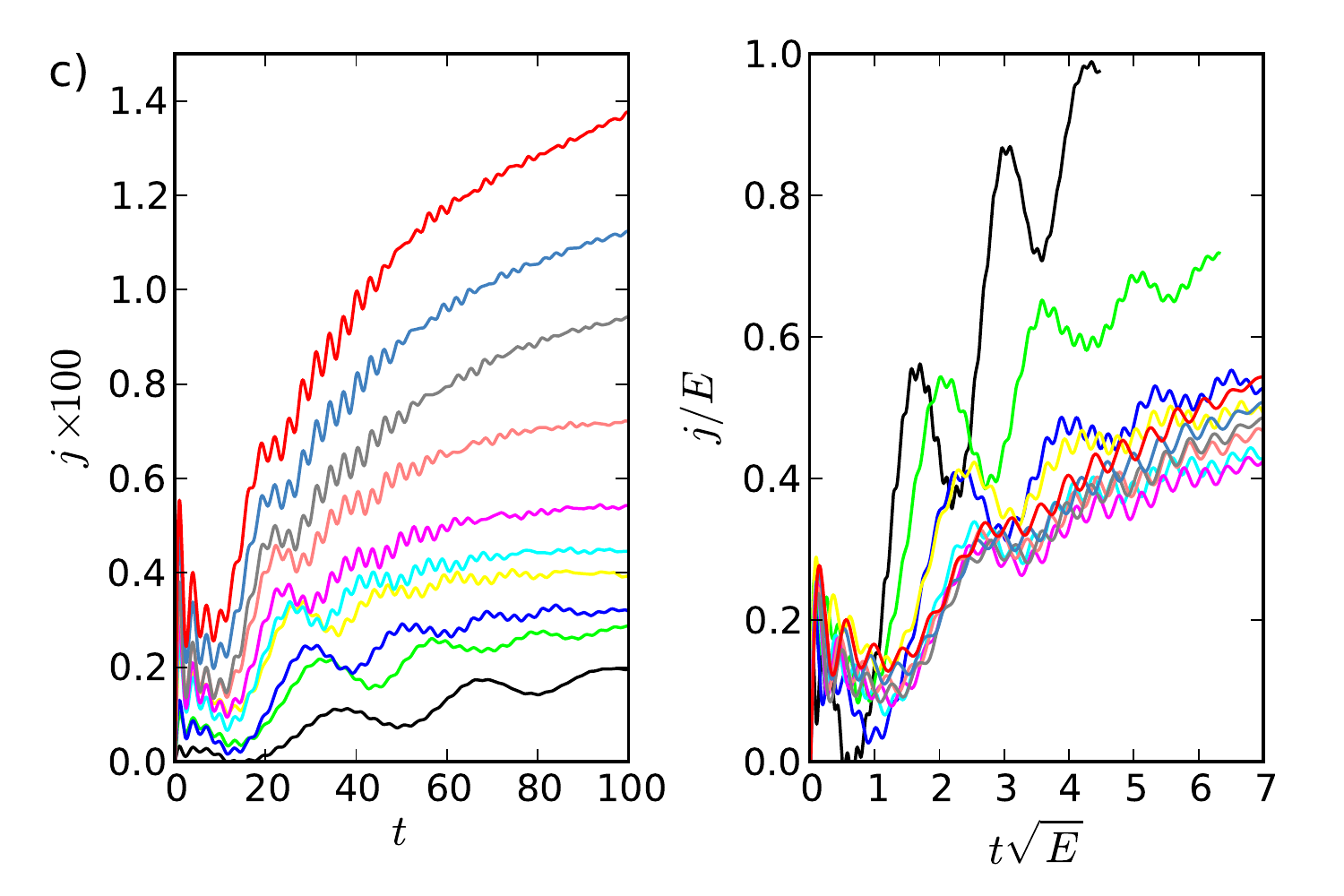}%
\includegraphics[width=0.29\textwidth]{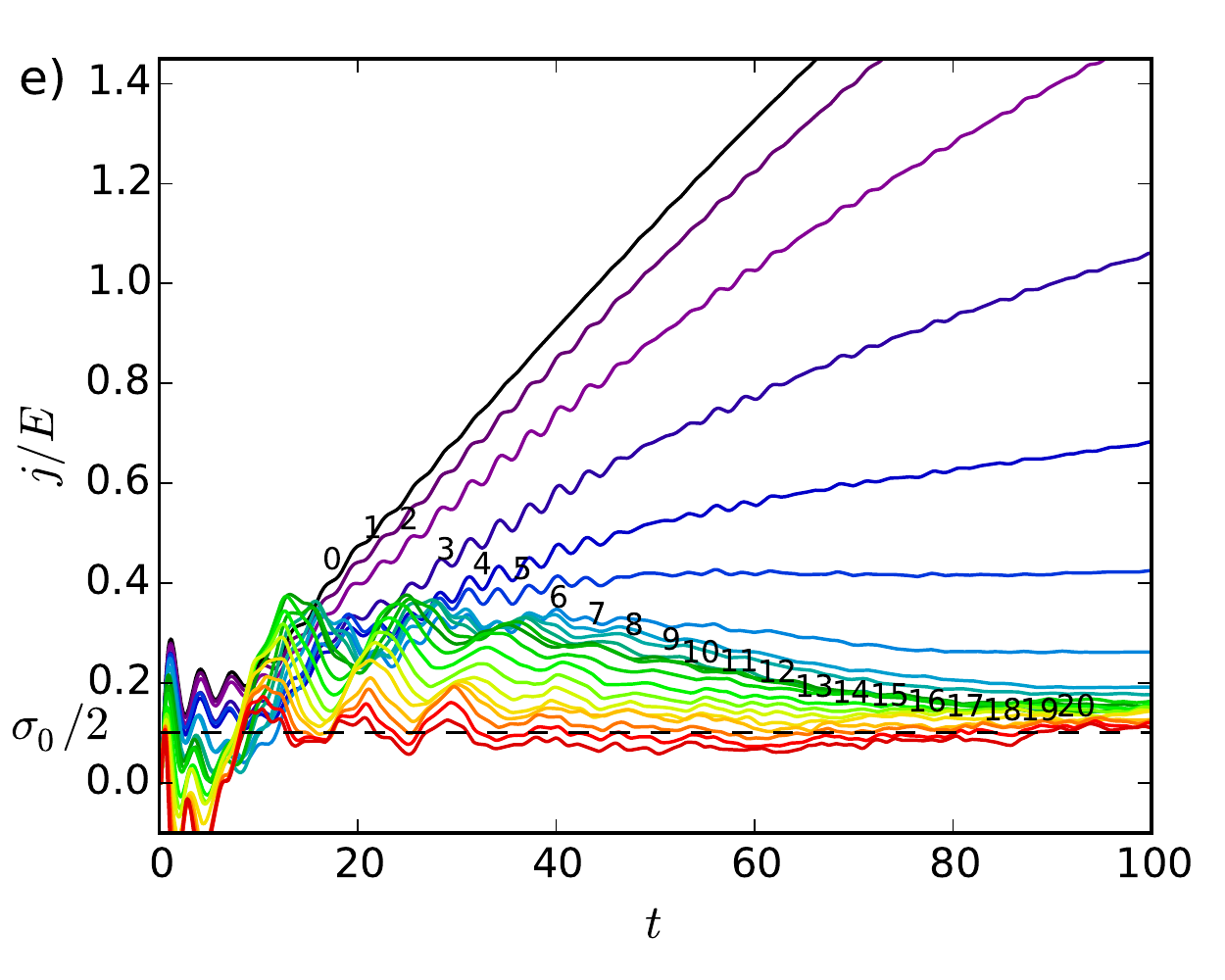}\\
\includegraphics[width=0.35\textwidth]{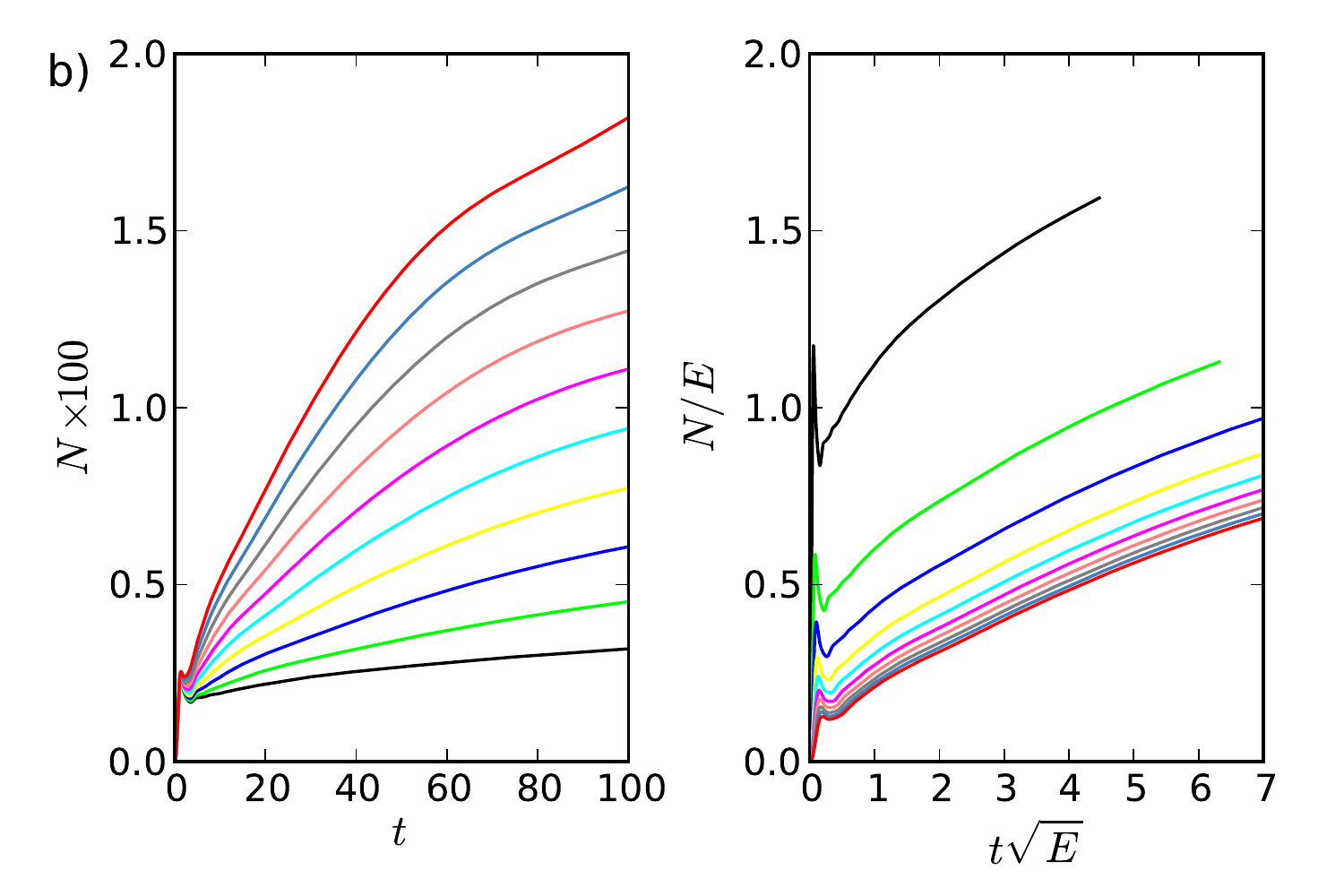}%
\includegraphics[width=0.35\textwidth]{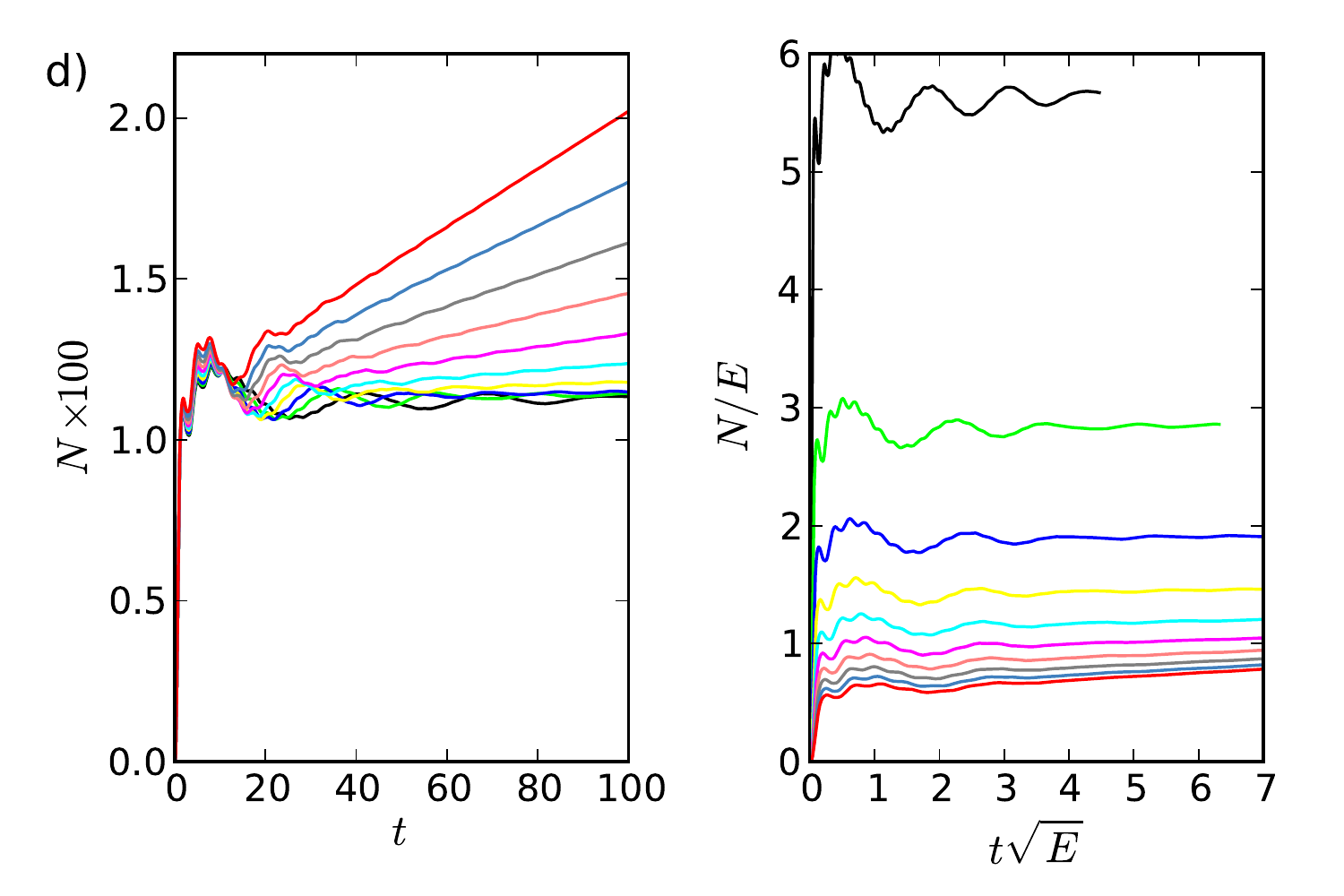}%
\includegraphics[width=0.29\textwidth]{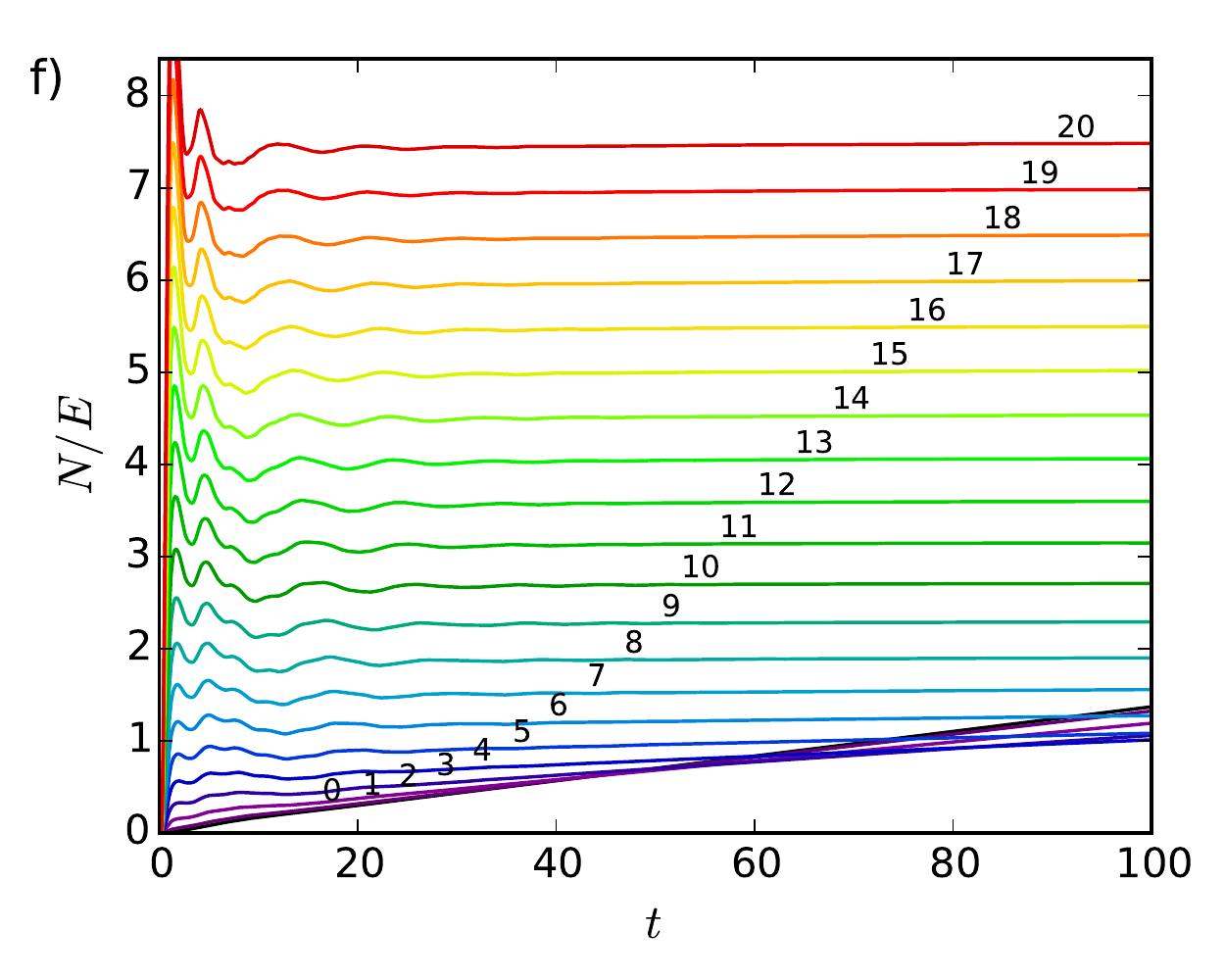}
\caption{\label{f:7} (color online). Dependence of the current (a,c,e) and pair creation rate (b,d,f) on the electric field and disorder. (a,b) Paramagnetic case, and (c,d) magnetic case for different values of the electric field (\(E=0.002\, E_0\), black, to \(0.02\, E_0\), red, as in Fig.~\protect\ref{f:2}), and \(J_I=0.2\, \nu\). Effect of the disorder for fixed electric field \(E=0.02\,E_0\), in the magnetic case: current (e), and pair creation rate (f); the lines correspond to 21 values \(J_I=0,\ldots, 1.0 \, \nu\) (from black to red, in steps of \(0.05\,\nu\), and numbered from 0 to 20). The dashed line in (e) shows that the current is above one half of the minimum conductivity \(\sigma_0/2\) for \(J_I \le 1.0\, \nu\). Averages are made over \(8192\) configurations of the impurities distribution. }
\end{figure*}

These qualitative changes in the electronic structure and in the phenomenology of the system's evolution, translate into a series of changes in the transport properties that become particularly important in the magnetic case. We show in Fig.~\ref{f:7} the time evolution of the current density and the pair production rate, in the paramagnetic (a,b) and magnetic (c,d) cases, for various values of the electric field (a-d), or of the disorder intensity in the magnetic state (e-f). To compare with the clean system of Fig.~\ref{f:3}, we also plot the scaled data [right panels in (a-d)]. It is worth noting that the initial evolution is strongly sensitive to the disorder configuration. We recall that the initial state is a hole located at the origin, and polarized with a spin up. Depending on the neighborhood, if it contains or not an impurity, or if the origin is occupied by an impurity, the individual evolution of the wave packet is different; this reflects by the existence of large statistical current fluctuations and in the pair production. This kind of dependency on the initial condition was already noted in the problem of two-dimensional quantum percolation \cite{Schubert-2008fk}. Therefore, the data corresponding to the weaker electric fields, \(E = 2,4,6 \times 10^{-3}\), did not completely converged after averaging over \(2^{13}\) configurations (black, green and blue lines in Fig.~\ref{f:7}). 

The nonlinear scaling behavior of the current \(j(t) \sim E^{3/2} t\), although preserved in the paramagnetic case (within the statistical errors), completely disappears in the magnetic case.  Even in the paramagnetic case and for weak disorder, there are differences with respect to the clean case: first, the initial pair creation rate jumps to a finite value, which in the range of electric fields used in the computations, appears to be independent of the electric field; second, in spite of the superposition of the scaled curves (right panels of Fig.~\ref{f:7}b), the characteristic straight line behavior as a function of time is much shorter than in the clean case. 

In comparison with the paramagnetic case, we note that the current traversing a magnetically polarized medium is reduced by a factor of about \(2\), for a given electric field (Fig.~\ref{f:7}c). Simultaneously and at first sight paradoxically, the number of pairs, and therefore the number of carriers, rapidly increases during an initial transient. Concomitantly, the spreading of the wave function is almost ballistic, in this weak to intermediate disorder strength regime (cf.\ Fig.~\ref{f:6}, for \(J_I<0.5\)). These observations show that we are in the presence of a regime characterized by an initial rapid spreading of the density probability, to which the pair creation rate is proportional, in conjunction with a slow displacement of its mean value, which determines the current. In addition, the absence of a definite power law in the current-electric field characteristics, in particular for the weaker electric fields \(E \le 0.01\,E_0\) (cyan line), can be related to the behavior of the number of pairs that tends to saturate. Therefore, in the magnetic case, the current driven by polarization dominates over the pair production term, erasing the power law dependency on the electric field.

To study the influence of the magnetic disorder on the current and pair production, we fixed the electric field at \(E = 0.02\, E_0\), and varied \(J_I\) between the clean value \(J_I = 0\) to a moderated disorder strength \(J_I = 1.0\,\nu\), limit of the ballistic regime (Figs.~\ref{f:7}e and \ref{f:7}f). We note that for a disorder strength of about \(J_I \approx 0.25\) (line 5), the current and pair production rate tend to saturate to a constant value (independent of time), after an initial transient regime. Increasing the disorder the current does not vanish, but appears to converge (within the large fluctuation errors) to a constant independent of \(J_I\). It is important to recall that the initial state is always at zero energy, that is in the energy gap open by the magnetic impurities (cf.\ Fig.~\ref{f:4}). This asymptotic value depends on the electric field. 

The pair production rate appears to be less influenced by stochastic fluctuations. This is justified by the fact that the pair production is computed in the comoving frame (the one in which \(k_y-Et\) is constant), and then it is not sensitive to the phase of the wave function at variance with the current. After an initial transient \(N(t)\) saturates to a value proportional to the disorder amplitude. This is in sharp contrast with the Schwinger mechanism that would give a rate exponentially small in the energy gap; for strong magnetic disorder the production of electron states from the initial hole state is arguably due to scattering off impurities and not directly related to the electric field intensity (as can be observed in Fig.~\ref{f:7}d, where the initial approximated discontinuity in \(N(t)\) at \(t=0^+\) do not depend on \(E\)).

Therefore, for increasing magnetic disorder in the ballistic or superdiffussive regime (\(J_I<1\)), the current decreases at long times, but remains above a minimum value, half of the minimum conductivity of clean graphene, \(\sigma_0/2\). Simultaneously the number of pairs increases proportionally to the disorder strength.

\section{Local density of states and localization}
\label{s:ldos}

%
% FIG 8
\begin{figure*}
\centering
\includegraphics[width=0.28\textwidth]{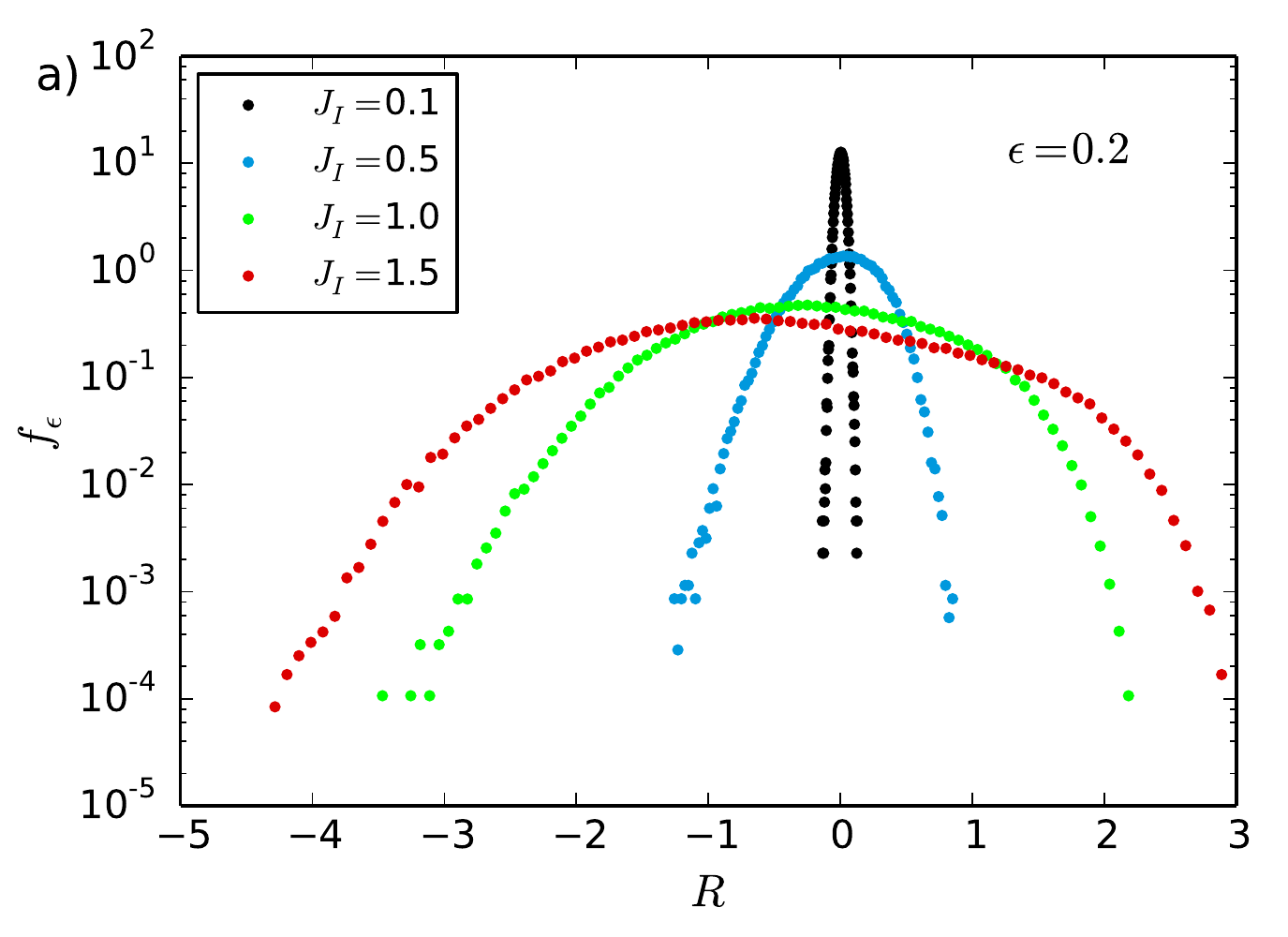}%
\includegraphics[width=0.35\textwidth]{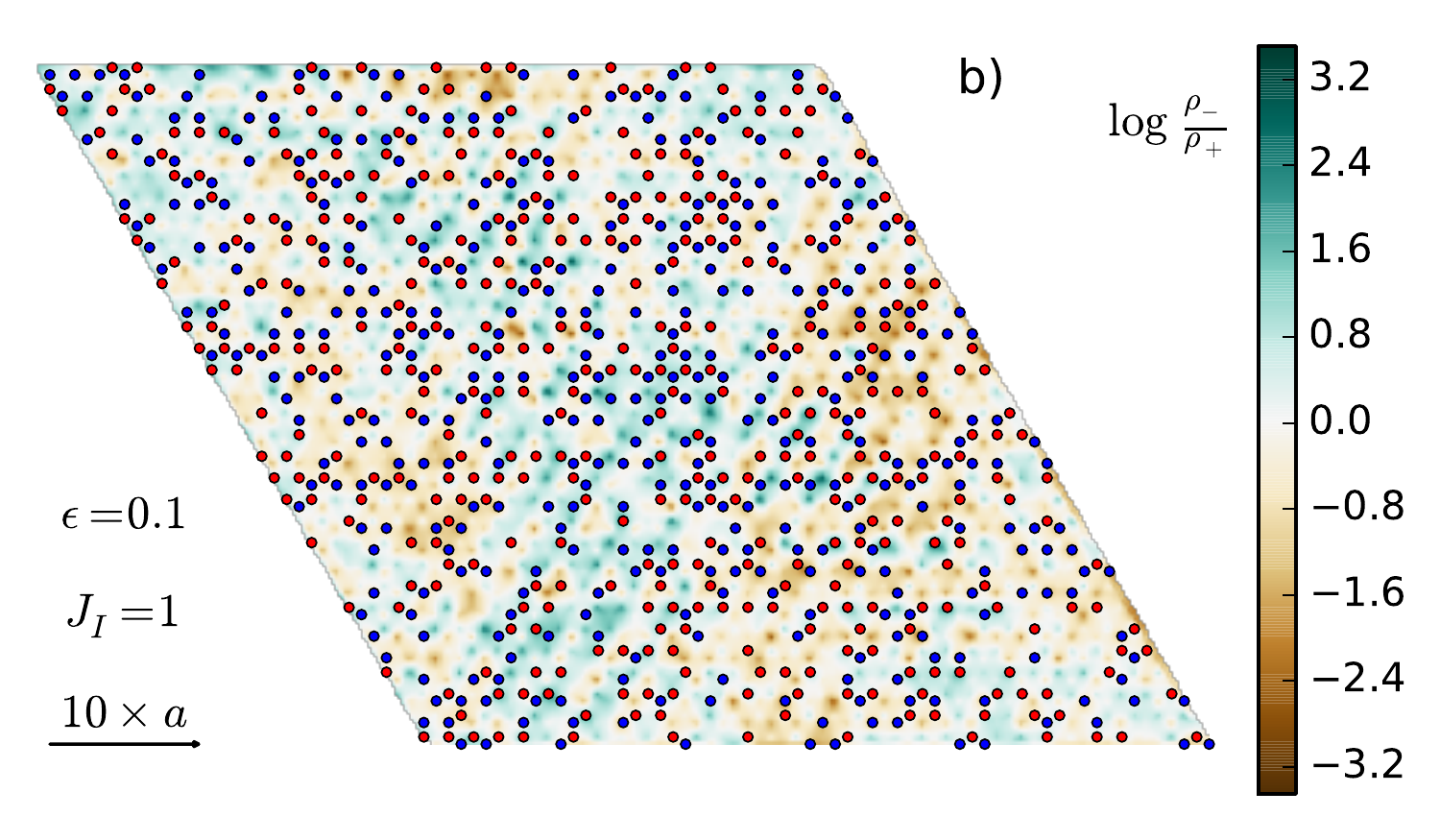}%
\includegraphics[width=0.35\textwidth]{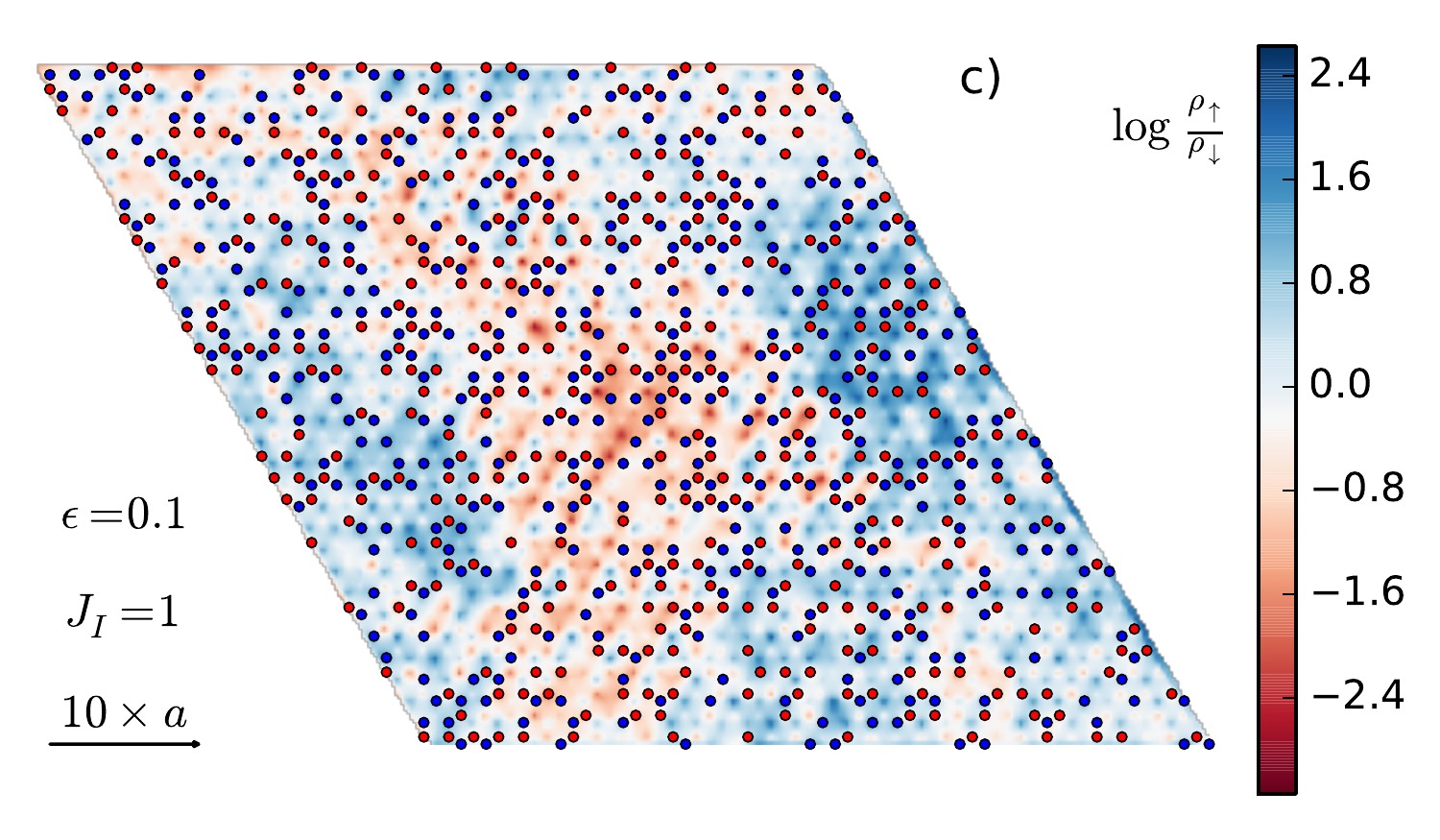}
\caption{\label{f:8} (color online). Histogram and spatial distribution of the local density of states in the magnetic case, showing a multifractal distribution that extends with increasing disorder strength, together with a strong spatial charge-spin correlation. (a) Energy \(\epsilon=0.2\,\nu\) near the gap edge for different values of \(J_I\); (b) electron \(\rho_+(i,\epsilon)\), hole \(\rho_-(i,\epsilon)\) distribution, and (c) spin up \(\rho_\uparrow(i,\epsilon)\), spin down \(\rho_\downarrow(i,\epsilon)\) distribution of states \(\epsilon=0.1\), for \(J_I=1.0\,\nu\). The hole patches are predominantly spin down, and electrons spin up, for the given disorder configuration. The circles locate the random impurities (spin up, blue; spin down, red). The histogram is averaged over \(2048\) sites times \(128\) disorder configurations. In (b-c), we show a region of \(2\times 32^2\) sites. }
\end{figure*}

The observed complex behavior of the wave packet and the peculiar properties of the current in the presence of polarized magnetic impurities, cannot be simply explained by the mechanisms of ballistic transport and electron-hole production in a strong electric field, suitable for the paramagnetic case. The fact that a gap is open and that a current weakly dependent on the exchange constant for strong enough disorder persists at long times, are indicative of interesting localization properties and highly inhomogeneous electronic states. More specifically, the current tends to a constant corresponding to half the clean minimum conductance, that can be a consequence of a spin dependent scattering and selective localization: one of the spin species eventually ceases to contribute to the charge transport.

This behavior, resulting from the interaction of the itinerant spins and the magnetic moments of the impurities, can be investigated using the local density of states,
\begin{equation}
    \label{e:ldos}
    \rho(i,\epsilon) = \sum_n \langle n|c_i^\dag c_i | n \rangle 
    \delta(\epsilon-\epsilon_n)\,,
\end{equation}
where \(\epsilon_n\) is one eigenvalue of the Hamiltonian (\ref{e:H}) corresponding to the eigenvector \(|n\rangle\). In addition, the statistical properties of \(\rho(i,\epsilon)\) can be related to the localization and critical properties of the electronic states, and thus used to characterize the metal-insulator transition \cite{Mirlin-1994kx,Dobrosavljevic-2003kx}. The existence of localized or critical extended states is related to strong spatial fluctuations of \(\rho(i,\epsilon)\). The probability distribution of the local density of states change from normal to log-normal, and thus its mean value \(\langle \rho_i \rangle\), which coincides with the density of states, differs from its typical, geometric mean value \(\exp\langle \log\rho_i \rangle\) \cite{Schubert-2010kx}. 

We show in Fig.~\ref{f:8} the histogram \(f_\epsilon[R]\) of the logarithm of the local density of states \(R=R(i,\epsilon)=\ln [\rho(i,\epsilon)/\rho(\epsilon)]\), at fixed energy \(\epsilon\), together with the spatial distribution of states resolved in energy \(\rho_\pm(i,\epsilon)\), and spin \(\rho_{\uparrow\downarrow}(i,\epsilon)\). The characteristic log-normal distribution of the local density of states, that should appear as an inverted parabola in Fig.~\ref{f:8}a, shift and deforms with increasing disorder strength. For weak disorder it is centered at the Fermi energy \(\epsilon=0\); for \(J_I=1.0\), near the transition between the superdiffusive to the subdiffusive regime, the peak of the distribution is in the low density side, showing a tendency to localization; for stronger disorder the states near the gap are localized (\(J_I=1.5\)). 

The most striking fact appears in the spatial distribution of electronic states shown in Fig.~\ref{f:8}b and c. The peculiar conductivity properties of graphene near the Dirac point measured in experiments \cite{Novoselov-2005kx}, were successfully related to the existence of large-scale charge inhomogeneities \cite{Martin-2008vn}. Electron-hole puddles were theoretically shown to arise in dirty graphene due to Coulomb (long range) impurities \cite{Adam-2007qf}, but can also form in the presence of short range impurities, as in hydrogenated graphene \cite{Schubert-2012vn}, or for other types of hybridation \cite{Garcia-2013vn}. The randomly distributed antiferromagnetic impurities break the translation invariance and sublattice symmetry (opening a gap), but preserving the electron-hole and spin symmetries. However, in the critical state (\(J_I\approx=1\)), we observe that large patches of separated electrons and holes are formed (Fig.~\ref{f:8}b), that are strongly correlated with a definite value of the carrier's spin (Fig.~\ref{f:8}c). Remarkably, the charged puddles are in fact spin polarized as in magnetic polarons \cite{Gennes-1960ud,Nagaev-2001fk,Rappoport-2011vn}. In this state, when an electric field is applied, we find that the conductivity is approximately \(\sigma_0/2\) (Fig.~\ref{f:7}e), a result compatible with the charge-spin selective scattering, which eliminates two of the four possible base states.

\section{Conclusion}
\label{s:conclusion}

We investigated the charge transport in graphene for two distinct cases of disorder. According to the magnetic polarization of impurities we distinguished the paramagnetic and the antiferromagnetic cases. The paramagnetic impurities create energy states around the Dirac point. Antiferromagnetic order of randomly distributed impurities, generates a gap proportional to the exchange coupling. A strong electric field, through the Schwinger mechanism, drives the production of electron-holes pairs and favors, in a disordered medium, an inhomogeneous charge polarization. 

The spreading of a wave packet follows a well defined power law in time, whose exponent depends on the disorder strength and type. In the paramagnetic case, increasing the disorder results in a smooth transition towards a diffusive regime. In the weak disorder range, the paramagnetic case is qualitatively similar to the clean case: the current depends nonlinearly on the electric field, with the characteristic exponent of the pair creation rate. At variance, in the antiferromagnetic case, a transition towards a subdiffusive regime occurs. We observed that even for relatively weak disorder, the pair creation is largely suppressed. The Schwinger mechanism, dominant in the paramagnetic case, is overwhelmed by charge polarization, and as a result, the linear response to the electric field is restored. However, while in the limit of weak disorder we measured a conductivity in agreement with the linear response of a clean system, for antiferromagnetic order, we found that it tends to half the clean value.

The superdiffusive to subdiffusive transition with increasing disorder, that takes place at a value where the hopping energy is of the same order as the exchange energy, is suggestive of localization effects. We considered this possibility by studying the local density of states. In the transition region, the distribution probability of the local density of states is log-normal, with a maximum shifted towards the low density region, implying the localization of the near gap states. These multifractal states are related to electron-rich and hole-rich patches, which in addition are spin polarized. The transport properties of the magnetic polaron state is characterized by a conductivity which is half the one of clean graphene; this is a consequence of the scattering on impurities that selects states with definite charge-spin correlation: electrons (positive energy) and holes (negative energy) patches acquire opposite spins and form a highly inhomogeneous texture.

\begin{acknowledgments}
We acknowledge helpful discussions with Laurent Raymond. Part of the computations were performed at the ``Mésocentre, Université d'Aix-Marseille.''
\end{acknowledgments}

%
%
%\bibliography{total}
%

\end{document}